\documentclass[usegraphicx,usenatbib,letterpaper]{mn2e}
\usepackage{graphicx}
\usepackage{myaasmacros}
\usepackage{mathrsfs}
\usepackage{amsmath}
\usepackage{amssymb}
\usepackage{amsfonts}
\usepackage{graphicx}
\usepackage{wasysym}
\usepackage{dblfloatfix}
\usepackage{color}
\usepackage{booktabs}
\usepackage{float}
\usepackage[normalem]{ulem}

\newcommand{\Ramses}{{\sc ramses}}
\newcommand{\RamsesRT}{{\sc ramses-rt}}
\newcommand{\pyFC}{{\sc pyFC}}

\newcommand{\Le}[1]{{\textit{L#1}}}
\newcommand{\smallC}{{\textit{smallC}}}
\newcommand{\medC}{{\textit{medC}}}
\newcommand{\bigC}{{\textit{bigC}}}

\newcommand{\twoFone}[1]{\ensuremath{{}_2F_1}}
\renewcommand{\vec}[1]{\ensuremath{\boldsymbol{#1}}}

\graphicspath{{Figures/}{Pictures/}}

\title[Feedback from Radiatively-driven AGN Winds]{
Outflows Driven by Quasars in High-Redshift Galaxies with Radiation Hydrodynamics}
\author[R. Bieri  et al. ]{
\parbox[t]{\textwidth}{
Rebekka Bieri$^{1}$\thanks{E-mail: bieri@iap.fr},
Yohan Dubois$^1$, Joakim Rosdahl$^{2}$, Alexander Wagner$^3$, \\
Joseph Silk$^{1,4,5,6}$, and Gary A. Mamon$^1$}
\vspace*{6pt} \\
$^{1}$ Institut d'Astrophysique de Paris (UMR 7095: CNRS \& UPMC -- Sorbonne
Universit\'es), 98 bis bd Arago, F-75014 Paris, France\\
$^2$ Leiden Observatory, Leiden University, P.O. Box 9513, 2300 RA, Leiden, The Netherlands \\ 
$^3$ Center for Computational Sciences, University of Tsukuba, 1-1-1 Tennodai, Tsukuba, Ibaraki, 305-8577, Japan \\ 
$^4$ Laboratoire AIM-Paris-Saclay, CEA/DSM/IRFU, CNRS, Univ. Paris VII, F-91191 Gif-sur-Yvette, France\\
$^5$ Department of Physics and Astronomy, The Johns Hopkins University Homewood Campus, Baltimore, MD 21218, USA\\
$^6$ BIPAC, Department of Physics, University of Oxford, Keble Road, Oxford OX1 3RH
}
\date{Accepted . Received ; in original form }

\begin{document}
\maketitle

\begin{abstract}
  The quasar mode of Active Galactic Nuclei (AGN) in the high-redshift Universe
  is routinely observed in gas-rich galaxies together with large-scale
  AGN-driven winds.  It is crucial to understand how photons emitted by the
  central AGN source couple to the ambient interstellar-medium to trigger
  large-scale outflows.  By means of radiation-hydrodynamical simulations of
  idealised galactic discs, we study the coupling of photons with the
  multiphase galactic gas, and how it varies with gas cloud sizes, and the
  radiation bands included in the simulations, which are ultraviolet (UV),
  optical, and infrared (IR).  We show how a quasar with a luminosity of
  $10^{46}\, \rm erg\, s^{-1}$ can drive large-scale winds with velocities of
  $10^2-10^3 \, \rm km \, s^{-1}$ and mass outflow rates around $10^3 \, \rm
  M_\odot\, yr^{-1}$ for times of order a few million years. Infrared radiation
  is necessary to efficiently transfer momentum to the gas via multi-scattering
  on dust in dense clouds.  However, IR multi-scattering, despite being
  extremely important at early times, quickly declines as the central gas cloud
  expands and breaks up, allowing the radiation to escape through low gas
  density channels. The typical number of multi-scattering events for an IR
  photon is only about a quarter of the mean optical depth from the center of
  the cloud.  Our models account  for the observed outflow rates of
  $\sim$~500-1000~M$_{\odot}\, \rm{yr}^{-1}$ and high velocities of $\sim
  10^3\, \rm km\, s^{-1}$, favouring winds that are energy-driven via extremely
  fast  nuclear  outflows,  interpreted here  as being  IR-radiatively-driven
  winds. 
\end{abstract}


%------------------------------------------------------------------------------------------------------------------
\section{Introduction}\label{sec:intro} %-----------------------------------------------------

Galaxy formation and evolution is highly non-linear and poses significant
challenges to our current understanding of the Universe.  One particular issue
is the relation between the mass function of dark matter halos, provided by
theoretical models, and the luminosity function of galaxies, given by
observations and usually fit by a \citet{Schechter1976} function.  By assuming
that stellar mass follows halo mass, we are left with a theoretical prediction
that leads to excessive numbers of  galaxies at both the low-mass and the high-mass ends.
Thus, some mechanisms have been advocated to regulate the baryon budget in
galaxies.  Feedback is likely an important mechanism, operating within galaxies
and driving large-scale outflows, removing the star-forming gas and/or
preventing further infall.  In high-mass galaxies, efficient feedback is
thought to be provided by active galactic nuclei (AGN) hosting supermassive
black holes (SMBHs) \citep[e.g.][]{Magorrian1998, Hu2008, Kormendy+2011}.  Gas
accretion onto a black hole (BH) leads to energy release capable of driving
outflows that regulate star formation and the local baryonic content
\citep{Silk+Rees1998}, hence regulating the BH growth and that of the
surrounding galaxy \citep{Kormendy+Ho2013}.

The two main modes of AGN feedback identified so far are the so-called
\textit{radio}-mode (radiatively-inefficient) powered by mechanical jets, and
the radiatively-efficient \textit{quasar}-mode powered by photons that couple
to the gas and transfer their momentum to it.  Several studies have examined
the effects of these different modes in cosmological hydrodynamical simulations
\citep[e.g.][]{DiMatteo+Springel+2005, DiMatteo+2008, Sijacki+2007,
BoothSchaye2009, duboisetal2012} and also on smaller galaxy scales
\citep[e.g.][]{Proga+2000, Ostriker+2010,
Wagner+Bicknell2011,Nayakshin+Zubovas2012, Gabor+Bournaud2014}, and have
highlighted the capacity of regulating the baryon content of galaxies with AGN
feedback.  The general findings of these hydrodynamical simulations and also of
semi-analytical models \citep[e.g.][]{Croton+2006,Bower+2006} is that AGN
feedback suppresses star formation in massive galaxies, reproduces the observed
high-end tail of the galaxy mass function, and is largely responsible for the
morphological transformation of massive galaxies into ellipticals.

Although the picture of supermassive black holes exerting strong feedback on
their host galaxies is very attractive, the details of the mechanism remain
vague, primarily because the implementation of the black hole feedback in these
studies relies on subgrid recipes in pure hydrodynamical (HD) simulations. The
implementations in these studies vary slightly, but in most cases, quasar
feedback is approximated by depositing thermal energy within the resolution
element, with the efficiency of the radiation-gas coupling represented by a
single parameter chosen to match global observations of SMBH mass-bulge
velocity dispersion~\citep{Ferrarese+Merritt2000}.  This is the quasar mode of
AGN feedback implemented in recent $\sim 100\, \rm Mpc$ state-of-the-art
hydrodynamical cosmological simulations such as Horizon-AGN~\citep{HAGN},
Illustris~\citep{Illustris}, MassiveBlack-II~\citep{MBII}, or
Eagle~\citep{Eagle}.

In reality, this sub-grid model of quasar feedback hides complex physics. There
is a general consensus that photons emitted from a thin accretion disc
surrounding the SMBH~\citep{Shakura&Sunyaev1973} will effectively couple to the
surrounding interstellar medium (ISM) and eventually drive a large-scale wind.
These radiatively-driven winds are not necessarily simply momentum-conserving
with $\dot p \simeq L/c$ (as proposed by~\citealp{King2003}), where $\dot p$ is
the momentum input rate, $L$ is the bolometric luminosity of the source, and
$c$ is the speed of light, as opposed to energy-conserving, where the momentum
of the gas builds up from the pressure work of the gas, hence, $\dot p \gg
L/c$~\citep{Silk+Rees1998, FGQuasar2012, Zubovas&King2012}.  For example,
observations by \citet{Cicone+2014} show outflows with mechanical advantages of
a few tens, where the mechanical advantage is defined as the ratio between
$\dot p$ and $L/c$.  Recent observations of fast ($>10^3 \,\rm km\,s^{-1}$)
molecular outflows from AGN favour energy-conserving winds from the nuclear
accretion disc \citep{Tombesi+2015, Fergulio+2015} as do recent multiphase
simulations discussed below, \citep[e.g.][]{Costa+2014}.

Alternate approaches to simple internal energy input have been implemented,
where the gas close to the BH is explicitly given a momentum input rate of
$L/c$ multiplied by a factor of a few~\citep{DeBuhr+2010, DeBuhr+2011,
Choi+2012, Choi+2014,Barai+2014, Zubovas&Nayakshin2014, Costa+2014,
Hopkins+2016}.  However, simulations show contradictory results regarding the
plausibility of regulating the BH growth and impact on the baryon content of
the galaxy using momentum-driven winds.  While~\citet[][see
also~\citealp{Barai+2014}]{Costa+2014} advocate an energy-conserving wind to
significantly affect the galaxy and its surroundings,~\cite{Choi+2012} show
that their momentum feedback drives faster winds than their energy feedback
model.  Since quasar-driven winds are powered by the complex coupling of
photons with the gas, it becomes timely to improve our current understanding of
the transfer of momentum and energy from AGN-emitted radiation to the ISM by
means of radiation-hydrodynamical (RHD) simulations. In such RHD simulations,
the emission, absorption, and propagation of photons and their interaction with
the gas is self-consistently followed.  Detailed RHD simulations should help to
improve our understanding in how the quasar radiation couples to the gas, in
particular via the coupling of infrared (IR) radiation to dust, and how
powerful quasar winds are eventually driven.  Therefore, RHD simulations can
provide better sub-grid models for large-scale cosmological simulations.

RHD simulations of quasar feedback on galactic scales have become feasible
\citep[e.g.,][]{Ciotti+Ostriker2007, Ciotti+Ostriker2012, Kim+2011, Novak+2012}
and allow us to measure how much of the radiation momentum is transferred to
the gas.  These studies manage to resolve important physics involving the
variability of the AGN (the \emph{duty cycle}). RHD simulations also show how a
strong radiation source at the centre of a galaxy affects the surrounding ISM.
\cite{Ciotti+Ostriker2007, Ciotti+Ostriker2012, Novak+2012} show that, for
elliptical gas-poor galaxies, the momentum input rate very rarely exceeds $L/c$
due to the low opacity of dust to the re-radiated IR and the destruction of
dust in high temperature environments, which in turn limits the amount of
momentum the radiation can transfer to the gas.  \cite{Kim+2011} simulate the
evolution of a high-redshift galaxy with radiation from the SMBH, however, the
IR radiation has not been considered in this work, and, hence, the impact of
radiation into the gas is probably underestimated. Similarly
in~\cite{roosetal}, the effect of the UV photo-heating only from a central
quasar (treated in post-processing) has very little impact on the evolution of
a gas-rich isolated disc galaxy.  The IR radiation is potentially important
because in optically-thick gas it is constantly absorbed by dust and
re-emitted, and hence can give several times its momentum $L_{\rm IR}/c$ to the
dust up to the dust optical depth $\tau_{\rm IR}=\Sigma_{\rm g} \kappa_{\rm
IR}$, where $\Sigma_{\rm g}$ is the gas surface density and $\kappa_{\rm IR}$
is the dust opacity.  It can create radiation pressure-driven winds where the
momentum input rate exceeds the momentum flux of the source by one or two
orders of magnitude , mimicking the effect of an energy-driven wind.  Star
clusters in starburst galaxies can easily reach values of $\tau_{\rm
IR}=10-100$ (see Fig.~3 of~\citealp{Agertz+2013}), possibly explaining the
observed energy-conserving winds in high-redshift quasars.

High-resolution hydrodynamical simulations of AGN jet feedback
\citep[e.g.,][]{Bicknell+Sutherland+2000, Sutherland+Bicknell2007,
Antonuccio-Delogu+Silk2010, Wagner+Bicknell2011, Gaibler+2012} have shown that
a clumpy interstellar structure results in interactions between the jet and the
gas that differ from those in simulations with a homogeneous ISM. It is hence
expected that multiphase ISM properties also affect the momentum transfer from
the radiation to the gas. Modelling a clumpy interstellar structure might be
even more important when considering photons, as low density gas can trace
escape paths in which the radiation-matter interaction is reduced because the
radiation field is not fully trapped.  Also, it is expected that as the
radiation is sweeping up the dense gas, low-density channels form and photons
start to escape along these preferred directions, lowering the mechanical
advantage~\citep{KT2012,KT2013,Davis+2014,RT2015}.

The aim of this paper is to quantify the coupling of quasar radiation with a
clumpy ISM and to study how photons (UV, optical and IR) can drive powerful
winds using RHD simulations.  Since radiation-gas coupling depends on the
clumpiness of the gas, we consider a two-phase (see
\citealp{McKee+Ostriker2007}) fractal interstellar medium at pc-scale
resolution. We investigate the momentum budget associated with the dispersion
of the clouds in the galaxy and its dependence on different cloud sizes,
filling factors, quasar luminosities, and energy bands of
the quasar spectrum.

In Section~\ref{sec:setup}, we describe our suite of RHD
simulations. Our results are presented in Section~\ref{sec:results}.
Section~\ref{sec:discussion} is a discussion of the caveats of our
methods and setup.  Section~\ref{sec:conclusion} provides the final
conclusions.

% ------------------------------------------------------------------------------------------------------------------
\section{Methods}\label{sec:setup} 
% --------------------------------------------------------------------
We perform a suite of simulations using \RamsesRT{}, an RHD extension
of the adaptive mesh refinement (AMR) code \Ramses{}
\citep{Teyssier2002}. We model quasar-emitted radiation interacting
with a surrounding multiphase ISM, in order to study how efficiently
radiation couples to the gas within the galaxy.

\subsection{Initial Gas Density Distribution}
\label{subsec:DensityDist}
We set up a gaseous disc with a two-phase ISM in pressure equilibrium, with a
uniform hot phase with temperature $T\sim 10^6\,\rm K$, and a cold $T\sim 1 -
10^4\,\rm K$ phase that is uniform on large scales, but very clumpy on small
scales.  Our ISM setup is described in detail in \citet{Wagner+2013},
which followed work done by \citet{Sutherland+Bicknell2007} investigating the
interaction of an AGN jet with a non-uniform ISM.

The density field for the cold phase is homogeneous on large scales, i.e. there
is no radial density gradient, but it is very clumpy on small scales: it
follows a single-point log-normal distribution and two-point fractal
statistics.  In a log-normal distribution, the logarithm of the ISM density is
a Gaussian, with mean $m$ and variance $s^2$. With $P \left( \rho \right)$
being the log-normal probability distribution of the mass density $\rho$, one
can write
\begin{equation}
P \left( \rho \right) = \frac{1}{\sqrt{2 \pi}\,s\, \rho} \exp \left[- \frac{\left( \ln \rho - m \right)^2}{2s^2}\right]\ ,
\label{eq:logNorm}
\end{equation}
with 
\begin{equation}
m = \ln \left( \frac{\mu ^2}{\sqrt{\sigma ^2 + \mu ^2}} \right), \quad s = \sqrt{\ln \left( \frac{\sigma ^2}{\mu ^2} + 1 \right)}\, ,
\label{eq:mv}
\end{equation}
where $\mu$ and $\sigma ^2$ are the mean and the variance of the linear density
field.  In the simulations presented here, we adopt $\mu = 1$ and $\sigma^2 =
5$, identically to what has been used by \citet{Wagner+Bicknell2011}. These
values are in agreement with ranges found by \citet{Fischera+2003} and
\citet{Fischera+Dopita2004}, observing the column density distribution in a
turbulent ISM.  The variance in Eq.~(\ref{eq:mv}) gives a measure of how
concentrated the mass is within the density cores, or, conversely, the fraction
of the volume within the low density regions.  With these values, gas densities
below the mean $\mu$ encompass one-quarter of the mass and occupy
three-quarters of the volume of the simulated disc. For further discussion of
the adopted values, we refer to \citet{Bicknell+Sutherland+2000}.

We define $F \left( \textbf{k} \right)$ to be the Fourier transform of the
density $\rho (\textbf{r})$, where $\textbf{k}$ and $\textbf{r}$ are the
wave-number and position vectors, respectively. The two-point structure of a
homogeneous turbulent medium is characterised in Fourier space by an isotropic
power spectrum $D (k)$ defined as
\begin{equation}
D (k) = \int k^2 F ( \textbf{k} ) F^* (\textbf{k}) \textrm{d} \Omega.
\label{eq:powerLaw} 
\end{equation}
The power-spectrum is proportional to a power-law with index $-5/3$ in order to
reproduce the spectrum driven by Kolmogorov turbulence.  The Fourier transform
of the density $\rho$ is proportional to the turbulent field usually described
by the velocity vector \citep{Warhaft2000}.  We follow
\citet{Wagner+Bicknell2011} and adopt a standard Kolmogorov power spectrum for
our non-uniformly distributed gas within the disc.

We want to stress that our initial setup is stationary and hence does not
capture the actual ISM turbulence (see \citealp{Kritsuk+2011} and references
therein).  We rather parametrise the non-uniform properties of a generic
turbulent medium and focus on characteristics such as the variance of the gas
density $\sigma ^2$ and the two-point self-similar power-law structures, and
hence rely on a range of previous experimental and theoretical results from
the field of turbulence. The initial distribution of the ISM that we adopt
should therefore be regarded as a physically motivated generalisation of a
inhomogeneous ISM, while it may not necessarily represent an accurate model of
a turbulent ISM.

The density distribution is set up via an iterative process, described first by
\citet{Lewis+Austin2002}, where we have adapted the \pyFC\ package
\citep{Wagner+Bicknell2011} to our needs (mainly parallelised part of the code
in order to have a smaller minimum sampling wave-number). Our density field
simultaneously follows log-normal single-point statistics (see
Eq.~[\ref{eq:logNorm}]) and a Kolmogorov power-law self-similar structure in
wave-number with index $-5/3$ and minimum sampling wave-number $k_{\rm min}$.
In real space, the minimum sampling wave-number determines the scale of the
largest fractal structure in the cube relative to the size of the cube.
Effectively, it is the average number of clouds per dimension divided by two.
For example, in one setup of our simulation, we used $k_{\rm min} = 5\,\rm
kpc^{-1}$ for a cube mapped to a disc with a diameter of $d = 3$~kpc. Then the
largest structures (clouds) extend to $R_{\rm c, max} = 3/(2\, k_{\rm min}) =
300$~pc.

Finally, we place the cube into the \Ramses\ simulation domain by reading each
density value within the cube and placing it into a cell of the \Ramses\ grid.
Here the resolution of the generated cube does not necessarily need to be the
resolution of the smallest cell within the \Ramses\ simulation box. In order to
obtain a cylindrical shape resembling a galaxy, the density cube is filtered,
in the $xy$-plane, by a symmetric flat mean density profile with mean cold
phase density $\langle n_w \rangle$ and radius $r = 1.5 \, \rm kpc$, and in the
$z$-plane the density cube is filtered by a step function with height $h = 0.3
\, \rm kpc$.  The porosity of the ISM arises by imposing a temperature roof
$T_{\rm roof}$ above which the gas is defined to be in the hot phase.  The mean
density of the cold ISM phase is chosen such that the total mass of the cold
phase is $\sim 2 \times 10^{10} \, \rm M_{\odot}$ and is around $500 \, \rm H
\, cm^{-3}$. The roof temperature is around $50 \, \rm K$ for the different
simulations. 

In our simulations, clouds are initially in pressure equilibrium with the
surrounding hot phase, where the pressure is set to be $P \sim 7 \times
10^{-12} \, \rm Pa$.  For the hot phase, whose temperature is fixed at $T_{\rm
h} \sim 10^7\, \rm K$, the density is constant and set as $n _{\rm H, h} = 0.01
\, \rm H \, cm^{-3}$ for all the simulations, while the initial metallicity is
set to zero. The cold gas is initialised with solar metallicity.  With these
values and with a disc radius of $1.5 \, \rm kpc$ and thickness of $0.15 \, \rm
kpc$, we simulate a typical compact, gas-rich, high-redshift
galaxy~\citep[e.g.][]{Tacconi+2010,Daddi+2010,Genzel+2010}.  

The density and pressure profiles of the hydrostatic environment in a massive
gas-rich proto-galaxy are fairly flat under the gravitational influence of the
bulge and dark matter halo (assuming a \citealp{NFW1996} profile;
\citealp[e.g.,][]{Capelo+2010}). Hence this justifies the uniform hot phase
distribution adopted in our simulation.  

The filling factor of cold phase within the disc is given by
\begin{equation}
f_V = {V_{\rm cold} \over V_{\rm tot}} \ ,
\end{equation}
where $V_{\rm cold}$ is the volume of the cold phase and $V_{\rm tot}$ the
total cylindrical volume of the region in which the density is distributed.

Since the minimum sampling wave-number relates to the largest fractal structure
in the cube, we will hereafter respectively refer to the simulations with
$k_{\rm{min}} = 30$~kpc$^{-1}$, $5\,\rm kpc^{-1}$, and $1\,\rm kpc^{-1}$ as
\smallC\ (smallest clouds), \medC\ (medium clouds), and \bigC\ (biggest
clouds). For our simulations we chose a mean cold phase density $\langle n_w
\rangle$ of $508 \, \rm H \, cm^{-3}$, $503 \, \rm H \, cm^{-3}$, and $435 \,
\rm H \, cm^{-3}$ for the \smallC, \medC, and \bigC\ simulations, respectively.
Additionally, we chose roof temperatures of $43 \, \rm K$, $70 \, \rm K$, and
$70 \, \rm K$ for the \smallC, \medC, and \bigC\ simulations, respectively. 

\subsection{Radiation Hydrodynamics}
\label{rhd}

Among the possible RHD implementations that are both helpful for
cosmological and galaxy-scale simulations
\citep[e.g.,][]{Petkova+Springel2009, Krumholz+Klein+McKee2011,
  Pawlik+Schaye2011, Wise+Abel2011, Jiang+Stone+Davis2012,
  Skinner+OstrikerE2013}, we chose \RamsesRT\ \citep{Rosdahl+2013,
  Rosdahl+2015}, implemented in the \Ramses\ \citep{Teyssier2002} AMR
HD code, to model the interaction of radiation from the central black
hole with the galaxy's interstellar gas. The evolution of the gas is
computed using a second-order unsplit Godunov scheme for the Euler
equations. We use the HLLC Riemann solver~\citep{Toro+1994} with the
MinMod total variation diminishing scheme to reconstruct the
interpolated variables from their cell-centered values. The \RamsesRT\
RHD extension to \Ramses\ self-consistently adds the propagation of
photons and their on-the-fly interaction with hydrogen and helium via
photoionisation, heating, and momentum transfer, as well as their
interaction with dust particles via momentum transfer.  The advection
of photons between grid cells is implemented with a first-order moment method,
whereas the set of radiation transport equations is closed with the M1 relation
for the Eddington tensor \citep{Rosdahl+2013}. The M1 closure relation
\citep{Levermore1984} can establish and retain bulk directionality of photon
flows, and can to some degree model shadows behind opaque obstacles.

The radiation is split into different photon groups, defined by frequency
bands.  For each photon group, the radiation is described, in each grid cell,
by the radiation energy density (energy per unit volume) and the bulk radiation
flux (energy per unit area per unit time), which corresponds to the radiation
intensity integrated over all solid angles.  \RamsesRT\ solves the
non-equilibrium evolution of the ionization fractions of hydrogen and helium,
along with photon fluxes and the gas temperature in each grid cell.  For the
lowest-energy IR group, the photons can give momentum to the gas multiple times
via absorption and re-emission.  Higher-energy radiation groups are absorbed by
dust and re-emitted (conserving energy) into the IR group, but they can also
interact with hydrogen and helium via photoionisation.  We use a subgrid scheme
to account for the trapping of IR photons in regions where the mean free path
is smaller than the grid spacing.  This scheme recovers the proper asymptotic
limit in the radiation diffusion regime (see~\citealp{Rosdahl+2015} for a
detailed discussion).

Since the Courant condition imposes that the time-step duration (and therefore
the computational load) scales inversely with the speed of light $c$, we apply
the so-called reduced speed of light approximation (RSLA; see also
\citealp{Gnedin+Abel2001, Rosdahl+2013}).  The rationale for the RSLA is that
as long as the radiation travels faster then ionisation fronts, the results of
RHD simulations are more or less converged with respect to the (reduced) speed
of light.

However, IR radiation is not photo-ionising, so it is not obvious
whether a reduced speed of light produces converging results,
especially when IR trapping becomes important.  For our simulations,
we chose a reduced speed of light fraction $c_{\rm red} / c = 0.2$,
leading to $c_{\rm red} \sim 6 \times 10^4$~km$\,$s$^{-1}$.  We test
our chosen reduced speed of light by performing convergence tests in
Appendix~\ref{app:SOL} but leave a detailed discussion to a
forthcoming paper.

We implement outflow boundary conditions, such that any matter that leaves the
simulation volume is lost to the system.  We however choose a sufficiently
large simulation box size $L_{\rm box}=96\,\rm kpc$ to ensure negligible mass
loss.  The simulation employs a coarse grid of cell size $L_{\rm
box}/2^9=187\,\rm pc$ and allows up to 5 additional levels of refinement, so
that the smallest cell size is $L_{\rm box}/2^{14} = 5.9\,\rm pc$.  The
refinement is triggered with a quasi-Lagrangian criterion ensuring that if the
gas mass within a cell is larger than 80~M$_{\odot}$ a new refinement level is
triggered.  We ensure that all of the disc is maximally refined at the
beginning of the simulation, leading to $\sim 10^7$ cells initially at the
maximum level of refinement.  For convergence studies, we have performed lower
resolution runs with a spatial resolution of $\Delta x = 11.6$~pc using a
maximum of 6 levels of additional refinement.

The gas in our simulations follows the equation of state for an ideal
monoatomic gas with an adiabatic index of $\gamma = 5/3$.  To keep our setup as
simple as possible and only probe the effect of radiation-matter coupling, we
neglect gas cooling\footnote{Hence the only role of metals in our simulations
is to set the dust opacities, as shown by Eq.~(\ref{eq:opacity2})}, star
formation, feedback from stars, and gravitational forces, but these effects
will be studied in future work. We discuss those caveats in
Section~\ref{sec:discussion}.

\subsection{Modeling the Quasar}
\label{sec:Quasar}
\begin{figure}
\includegraphics[width=\columnwidth]{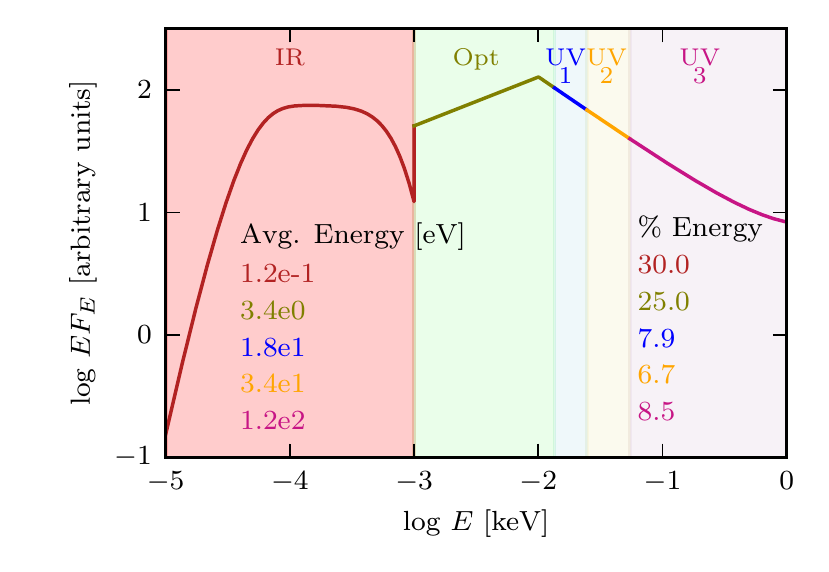}
\caption{Broad-band spectrum of a typical quasar (adopted from
  \citealp{Sazonov+2004}). The shaded areas show the energy range of
  each photon group: IR (red), optical (green), UV1 (blue), UV2
  (yellow), UV3 (pink). In the plot we show the fraction of the total
  energy going into each photon group (\% Energy) as well as the
  energy per photon (Avg. Energy) within the groups. Note that we do
  not model the hard X-ray energy band, which contributes 22\% to the
  total energy of the quasar.}
\label{fig:Spectrum}
\end{figure}
\begin{center}
\begin{table*}
\caption{Properties of the photon groups used in the simulations. Columns are
name; minimum and maximum energies; cross sections to ionisation by H\,{\sc i}, He\,{\sc i},
and He\,{\sc ii}; dust opacity; luminosity fraction (from quasar spectrum of
\citealp{Sazonov+2004}).  The dust opacity $\tilde{\kappa}$ for each group
scales with the gas metallicity $\kappa _i = \tilde{\kappa}_i Z/Z_{\odot}$.
The luminosity fraction per photon group is used to calculate the designated
energy from the quasar that goes into each corresponding photo group.} 
\begin{tabular}{llllllcc}
\specialrule{.12em}{.05em}{.05em}
Photon & \multicolumn{1}{c}{$E_{\rm min}$} & \multicolumn{1}{c}{$E_{\rm max}$}  &
\multicolumn{1}{c}{$\sigma_{\rm H\,\scriptscriptstyle I}$}  &
\multicolumn{1}{c}{$\sigma_{\rm He\,\scriptscriptstyle I}$} &
\multicolumn{1}{c}{$\sigma_{\rm He\,\scriptscriptstyle I}$} &
\multicolumn{1}{c}{$\tilde{\kappa}$} & 
\multicolumn{1}{c}{Energy} \\
group & 
\multicolumn{1}{c}{(eV)}   & 
\multicolumn{1}{c}{(eV)} &
\multicolumn{1}{c}{($\rm cm^2$)} & 
\multicolumn{1}{c}{($\rm cm^2$)} & 
\multicolumn{1}{c}{($\rm cm^2$)} & 
\multicolumn{1}{c}{($\rm cm^2\,g^{-1}$)} & 
\multicolumn{1}{c}{fraction}\\
\specialrule{.12em}{.05em}{.05em}
IR & \ \,$0.01$ & \ \,$1.$ & 0 & 0 & 0 & 10\ \, & 0.300   \\
Opt & \ \,$1.$ & $13.5$ & 0  & 0 & 0 & $10^3$ & 0.250   \\
UV1 & $13.5$ & $24.6$ & $3.1\!\times\!10^{-18}$ & 0 & 0 & $10^3$ & 0.079   \\
UV2 & $24.6$ & $54.4$ & $4.7\!\times\!10^{-19}$ & $4.2\!\times\!10^{-18}$ & 0
    & $10^3$& 0.067   \\
UV3 & $54.4$ & $10^3$ & $1.1\!\times\!10^{-20}$ 
    & $2.3\!\times\!10^{-19}$ & $1.7\!\times\!10^{-19}$ & $10^3$& 0.085   \\
\specialrule{.12em}{.05em}{.05em}	
\end{tabular}
\label{tab:rtinit}
\end{table*}
\end{center}
The spectrum of radiation from the central quasar is modelled using a maximum of
five photon groups, defined by the photon energy bands listed in
Table~\ref{tab:rtinit}.  The photons groups consist of one infrared (IR) group
(0.01~-~1~eV), one optical group (`Opt' from 1~-~13.5~eV), and three groups of
ionising ultraviolet (UV) photons, the first two (UV1 and UV2) bracketed by the
ionisation energies of H\,{\sc i}, He\,{\sc i} and He\,{\sc ii}, and the third (UV3) extending from
He\,{\sc ii} ionisation to soft X-rays (1~keV).  For a given quasar luminosity, we
split the spectral energy distribution (SED) of a typical quasar, as calculated
by \cite{Sazonov+2004} and shown in Fig.~\ref{fig:Spectrum}, into these 5
photon groups, and the corresponding fractions of the quasar luminosity are
given in Table~\ref{tab:rtinit}.  \citeauthor{Sazonov+2004} calculated the
typical quasar SED using published AGN
composite spectra in the optical, UV, and X-rays, also considering the cosmic
X-ray background and the contribution of AGN to infrared wavelengths, and the
estimated local mass density of SMBHs.  Note that we do not model the hard
X-ray energy band ($E > 1\,\rm keV$), where photons can heat (or cool) the gas
to an equilibrium temperature of $\sim2\times 10^7\,\rm K$ through Compton (or
inverse Compton) scattering of electrons, but this happens at a very small
distance from the source, and hence, these photons have little overall impact
on the ISM \citep[e.g.,][]{Ciotti+Ostriker2012, Hopkins+2016}.  The ionization
cross-sections $\sigma(E)$ are taken from \citet{Verner+1996}.  For each photon
group, the cross-sections are luminosity-weighted averages over the energy
interval, as described in \citet{Rosdahl+2013}.

Dust opacities are important for calculating the momentum transfer between the
photons and the gas as it happens via scattering on dust.  Interstellar dust
grains are destroyed, mostly by sputtering, by shock-heated gas above
temperatures of $T \ge 10^5$~K \citep{Draine+Salpeter1979a} (albeit the exact
temperature and destruction time scale depend on the dust grain size). For this
work, we modified the calculation of the dust opacity in \textsc{Ramses-RT} to
include dust destruction by thermal sputtering when the gas temperature is
above a cutoff temperature $T_{\rm cut}=10^5\,\rm K$.  This is necessary when
the source luminosity is sufficiently high to heat up the gas to very high
temperatures. The cut-off temperature for sputtering is weakly dependent on
density, an effect we ignore here.  We did not include dust sublimation since
\textsc{Ramses-RT} does not follow the dust temperature (which sublimes above
10$^3$~K) and leave this for future work.  

In our simulations the opacities for the different photon groups are given by 
\begin{equation}
\kappa_{i} = \tilde{\kappa}_{i} {Z \over Z_\odot} \,\exp\left(-{T \over T_{\rm cut}}\right) \ .
\label{eq:opacity2}
\end{equation} 
We thus assume that the dust content simply scales with the metallicity of the
gas below the cutoff temperature. 

For the IR ,we assume an opacity of
$\kappa_i \!= \! \kappa _{\rm IR} = 10 \, (Z / Z_\odot)\, \rm{cm}^2
\rm{g}^{-1}$,
whereas for the higher energy photons (optical and UV) we assume
$\kappa _i = 1000 \, (Z/ Z_\odot) \, \rm{cm}^2 \rm{g}^{-1}$, i.e.,
one hundred times higher than of the IR. The chosen opacities are
physically motivated by a combination of observations and
dust-formation theory of the ISM and stellar nurseries
(\citealp{Semenov+2003} for the IR photons, and
\citealp{Li+Draine2001} for higher energy radiation). They are however
uncertain by a factor of a few, due to model uncertainties as well as
the temperature dependence on the opacity, which is ignored in our
simulations. Past studies have used similar values
\citep[e.g.][]{Hopkins+2011, Agertz+2013, Roskar+2014}. The usual IR
opacities are in the range of
$\kappa _{\rm IR} = 5 - 10\, \rm{cm}^2 \rm{g}^{-1}$ and hence our
assumed IR opacity is at the high-end of what is usually considered.
 
Table~\ref{tab:rtinit} lists the values for the photon group energies, where
for each photon group, the energy intervals are between the lower bound $E_{\rm
min}$ and upper bound $E_{\rm max}$. Also shown are the photoionisation
cross-sections ($\sigma_{\rm H\,\scriptscriptstyle I}$, $\sigma_{\rm
He\,\scriptscriptstyle I}$, and $\sigma_{\rm He\,\scriptscriptstyle II}$) for
hydrogen and helium, calculated as described above, the dust-interaction
opacities ($\tilde{\kappa}_i$), and fractions of total quasar luminosity
emitted into each photon group.

We have chosen two different quasar luminosities $10^{43}$~erg$\,$s$^{-1}$ and
$10^{46}$~erg$\,$s$^{-1}$ for our simulations. The bolometric quasar luminosity
function (QLF) at redshift $z=3$, compiled by \citet{Hopkins+2007}, shows that
the chosen quasar luminosities are very common and nicely bracket the QLF.

In the present work, we explore the effect of an AGN radiation source
that is steady, isotropic, and located at the centre of the galaxy. In
reality, the quasar luminosity is proportional to its accretion rate,
which varies in time.  Additionally, the black hole can move around
the potential well of the dark matter halo and galaxy and hence is not
always exactly located at the centre of the galaxy.  Usually, a lower
(higher) density environment around the black hole results in a lower
(higher) luminosity of the quasar.  Moreover, changing the location of
the black hole changes the optical depth around the source, as the
encompassing density changes, which should alter the level of coupling
between the radiation and matter. For this reason, we have ensured
that the initial conditions for the density field are such that the
black hole for the $10^{46}$~erg$\,$s$^{-1}$ simulations is in the
densest region of the entire density distribution, while the initial
conditions place the lower luminosity source at an intermediate
density.  Aside from this, the statistical properties (as discussed
above) of the simulations with both quasar luminosities are exactly
the same.  We have also performed a simulation with a
$10^{46}$~erg$\,$s$^{-1}$ quasar surrounded by the exact same density
distribution as for the $10^{43}$~erg$\,$s$^{-1}$ simulations to
assess the role of quasar luminosity independently of the density
around the quasar.

\begin{figure*}
 \centering
 \leavevmode
 \includegraphics[width=\textwidth]{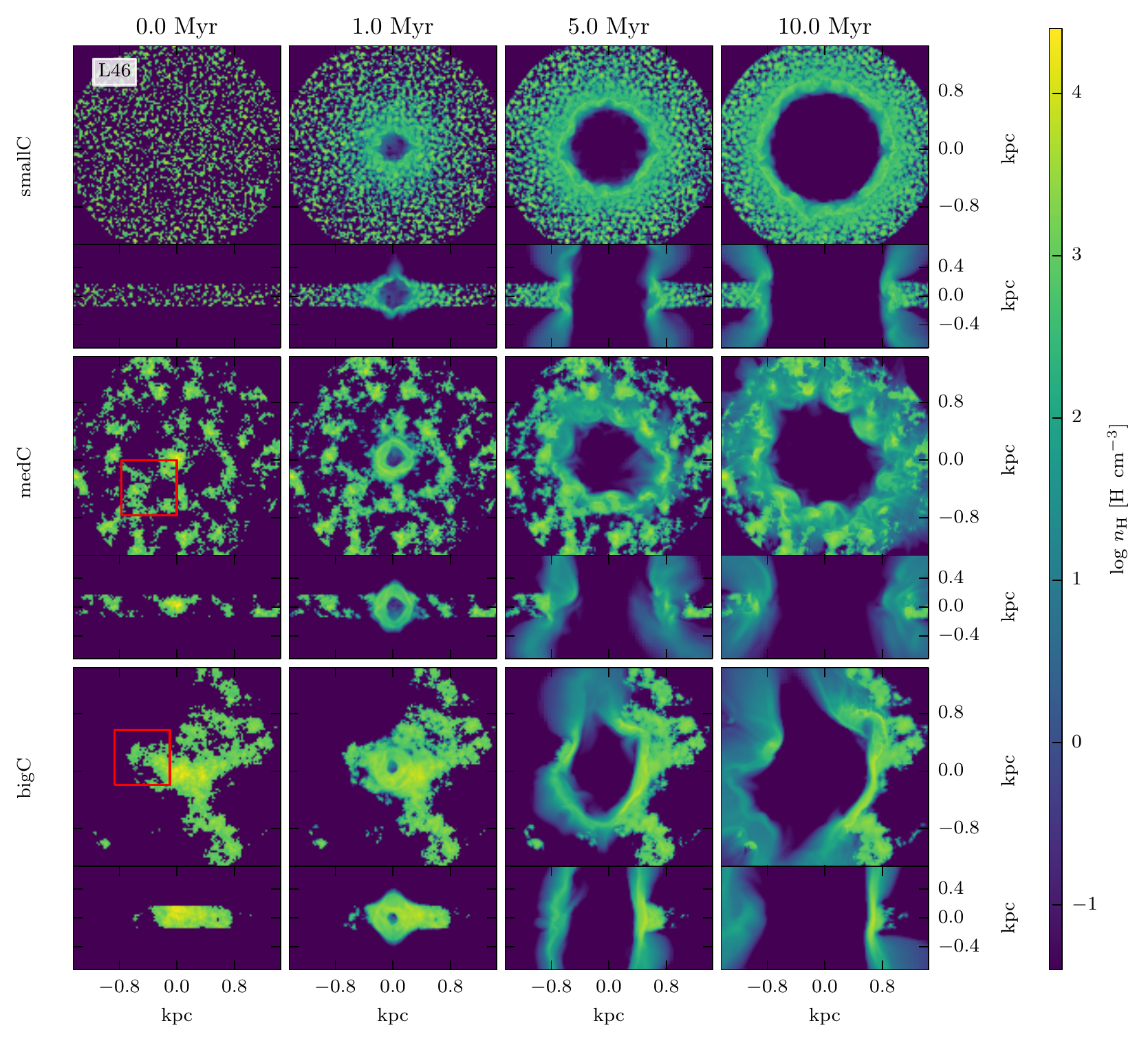}
 \caption{Slices of the gas density for the \Le{46}\_\smallC\ simulations (top row),
\Le{46}\_\medC\ simulations (middle row), and for the \Le{46}\_\bigC\
simulations (bottom row). The different columns show different times as
labeled.  The galaxies are shown both face-on (upper portion of rows) and
edge-on (bottom portion of rows). The galaxy is destroyed by radiation, and a
large-scale radiatively-driven wind is generated.  While in the \smallC\
simulation the wave induced by the radiation expands almost uniformly in the
$xy$-plane, this is not the case for the \medC\ and \bigC\ simulations where the
outflow escapes via tunnels of low density.  The red square over-plotted at the
0~Myr plot for the \bigC\ simulation corresponds to the zoomed-in region of
Figs.~\ref{fig:ZoomCloud} and \ref{fig:ZoomCloudPhotonGroups}, whereas the red
square in the \medC\ simulation corresponds to the zoomed-in region in
Fig.~\ref{fig:ZoomCloudIndividial}.}
\label{fig:Mape46DensfixedFill}
\end{figure*}

In summary, the key model parameters used in the different simulations are the
quasar luminosity, the radius of the largest fractal structure, and the
location of the quasar Table~\ref{tab:sim} summarises the assigned and derived
parameters that we used for our simulations and
Fig.~\ref{fig:Mape46DensfixedFill} shows an example of the \smallC, \medC, and
\bigC\ cloud distributions, where the volume filling factor is kept at 50\%. We
have also performed simulations with a filling factor of 100\%, but the results
do not significantly change from those shown in the paper.
\begin{table}
  \caption{Simulation parameters: quasar luminosity ($L$), largest possible
  cloud size ($R_{\rm c, max}$), and the gas density (`environment') around the
  quasar. If, for instance, the quasar environment is denoted with \textit{max}
  the quasar is placed into the cell within a maximum density.  Additionally,
  no suffix about the quasar position is used if the position of the quasar is
  in a maximum density environment for the \Le{46}\ simulation, or in a medium
  density environment for the \Le{43}\ simulation, respectively.} 
  \tabcolsep 3pt
\label{tab:sim}
\begin{center}
\begin{tabular}{l|ccl} 
\hline
Simulation  &$\log L$             &  $R_{\rm c, max}$  & Quasar \\
name        & (erg$\,$s$^{-1}$)   &  (kpc)             & environment \\ 
\hline
\Le{46}\_\smallC                    & 46 & 0.05 & \textit{max}  \\
\Le{46}\_\medC                      & 46 & 0.3  & \textit{max}  \\
\Le{46}\_\bigC                      & 46 & 1.5  & \textit{max}  \\
\Le{46}\_\bigC\_\textit{med$\rho_{\rm Q}$}      & 46 & 1.5  & \textit{med} \\ 
\Le{46}\_\bigC\_\textit{min$\rho_{\rm Q}$}      & 46 & 1.5  & \textit{min}  \\ 
\hline 
\Le{43}\_\smallC                    & 43 & 0.05 & \textit{med} \\
\Le{43}\_\medC                      & 43 & 0.3  & \textit{med} \\
\Le{43}\_\bigC                      & 43 & 1.5  & \textit{med} \\ 
\Le{43}\_\bigC\_\textit{max$\rho_{\rm Q}$}      & 43 & 1.5  & \textit{max} \\ \hline%
\end{tabular}
\end{center}
\end{table} 

%
% -------------------------------------------------------------------------
\section{Results}\label{sec:results} % -----------------------------------
We now present our simulation results and examine the interplay of the
BH-emitted radiation with the surrounding gas, focusing on the momentum
transferred from the radiation onto the gas.  We start with a comparison of the
effects of radiation on the galaxy for different cloud sizes for the more
luminous $L=10^{46}\,\rm erg\, s^{-1}$ source. Then we investigate how the wind
becomes radiatively-driven. We next compare the effects of different quasar
positions relative to the clouds, and end with a comparison of different quasar
luminosities. Additionally, we include in Appendix~\ref{app:SOL} a study of how
the transferred momentum depends on the speed of light and conclude that this
approximation is a crucial component in recovering the correct mechanical
advantage from the radiation onto the gas for the first few Myr.

\subsection{Effects of Different Cloud Sizes}
We explore the impact of radiation on the gas clouds and velocity evolution for
galaxy discs with different cloud sizes.  We focus here on the \Le{46}
simulations, where the source is embedded into the high density environment of
the clouds.

\subsubsection{Qualitative Effects of Cloud Sizes}
\begin{figure}
 \centering
 \leavevmode
 \includegraphics[width=0.93\columnwidth]{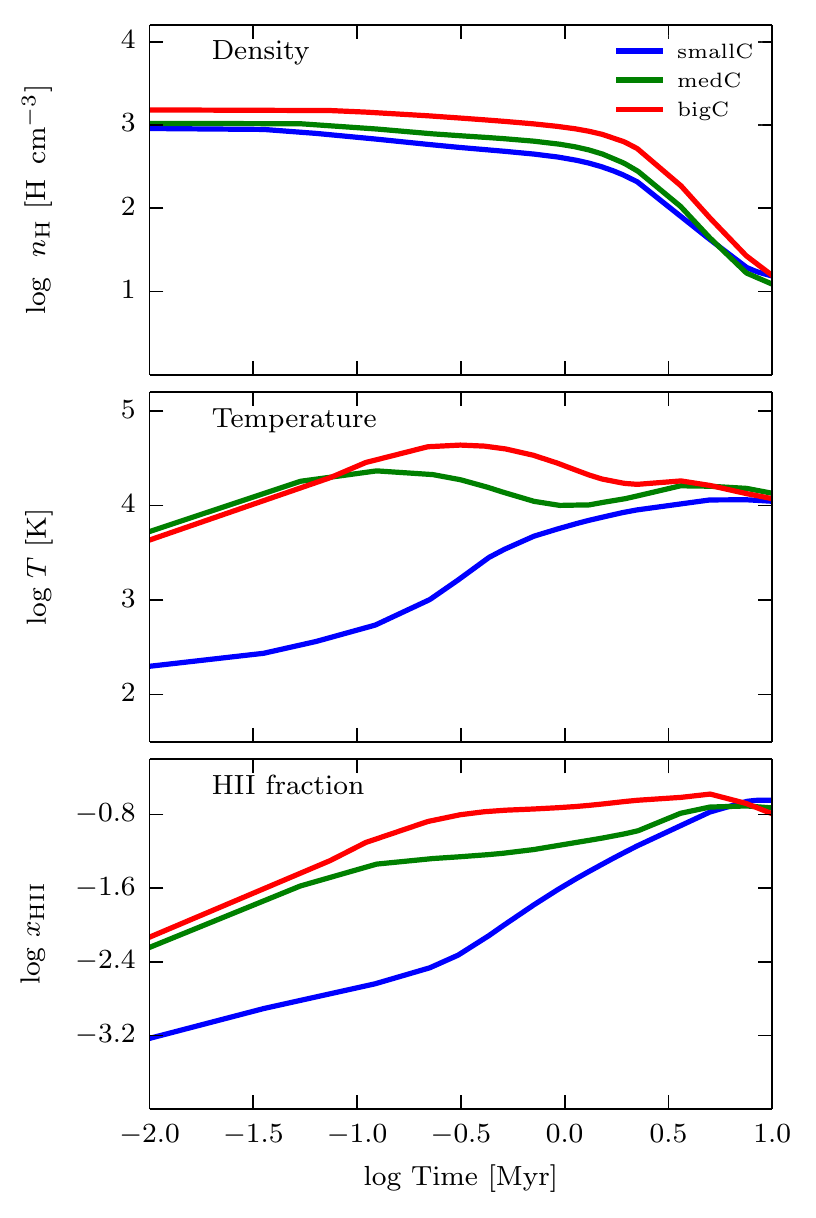}
 \caption{Evolution of the mean density (top), temperature (middle), and
H\,{\sc ii} fraction (bottom) of the clouds as a function of time for the
\Le{46}\_\smallC\, (blue), \Le{46}\_\medC\, (green), and \Le{46}\_\bigC\, (red)
simulations.  The lines show the mass-weighted mean value for the clouds, using
only the cells with a metallicity above a cutoff value of 0.5 Solar (where the
maximum and minimum metallicities are Solar and zero).  The radiation manages
to disperse the clouds, as can be seen by the decreasing mean densities. The
mean temperature as well as ionisation fraction increase with time. This lowers
the influence of the radiation as time passes due to the gas being either
transparent and/or ionised.} 
\label{fig:PropCloudLe46Max}
\end{figure}

In Fig.~\ref{fig:Mape46DensfixedFill}, we show maps of the gas density in the
disc at different times for the \Le{46}\_\smallC, \Le{46}\_\medC, and
\Le{46}\_\bigC\ simulations. Comparing the three different cloud sizes, we
observe that the outflow is less symmetric for less uniform initial conditions,
i.e.  larger clouds.  For all the simulations, the radiation from the central
source destroys the encompassing high density cloud hosting the source before
reaching a lower density environment. The ionised outflow generated by the
radiation pushes the gas out into the circum-galactic medium through the lower
density channels, which are more prominent with larger clouds.

Until the cloud is destroyed, the source is surrounded by a high density
environment where the optical depth is sufficiently high for the IR radiation
to boost the momentum transfer to the gas. Once the radiation manages to
destroy the central cloud, the optical depth drops to lower values, reducing
the effect of momentum-boost from IR multi-scattering. The time it takes for
the radiation to break from the central source through the cloud is hence
crucial and is much shorter for the \Le{46}\_\smallC\ simulation than for the
larger-cloud \Le{46}\_\medC\ or \Le{46}\_\bigC\ simulations.  This gives the
photons in the \Le{46}\_\medC\ and \Le{46}\_\bigC\ simulations more time to
transfer momentum to the gas via the trapped photons in the optically-thick
gas.

After the radiation has destroyed the encompassing cloud, the gas from the
cloud continues to expand until it reaches the neighbouring over-densities.
Since the clouds distributed within the \Le{46}\_\smallC\ simulation are all
relatively small, they are rapidly destroyed by the radiation, and the cloud
gas efficiently mixes with the background gas and fills up tunnels of lower
density, quickly creating smooth isotropic shells of outflowing gas.  Clouds in
the \Le{46}\_\medC, and \Le{46}\_\bigC\ simulations are much bigger, and, at
$t=1\,\rm Myr$, the shell evolution is less spherically-symmetric compared to
the \Le{46}\_\smallC\ simulation.  Additionally, the outflow generated by the
radiation creates a shell of swept-up gas, which is most apparent in the
\Le{46}\_\bigC\ and also at early times of the \Le{46}\_\medC\ simulation,
while less dominant for the \Le{46}\_\smallC\ simulation. Indeed, the larger
the clouds, the more gas mass they contain and the greater (and hence more
apparent) the overdensity of the shell can become.

\subsubsection{Cloud Evolution}
\label{subsec:Clouds}
\begin{figure*}
 \centering
 \leavevmode
 \includegraphics[width=0.93\textwidth]{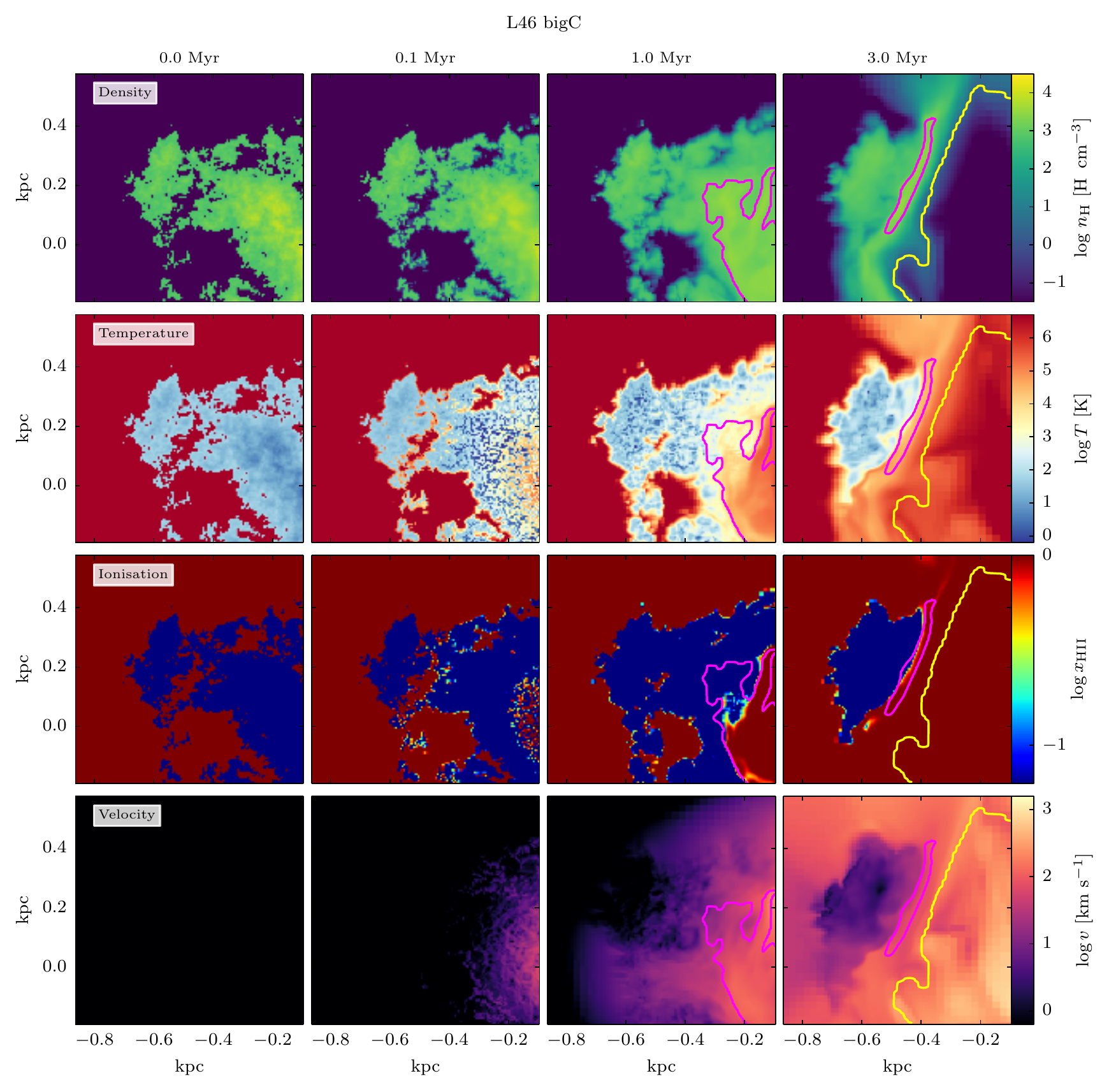}
 \caption{Slices of the gas density (top row), temperature (second to top row),
ionisation fraction (second to bottom row), and velocity field (bottom row) for
a slice of a zoomed region in the \Le{46}\_\bigC\ simulation. The position of
the zoomed-in region is marked by a red square in
Fig.~\ref{fig:Mape46DensfixedFill}.  To guide the eye, selected density
contours are overplotted at 1~Myr and 3~Myr. At 1~Myr, the magenta contour
denotes a density of $4000 \,\rm H\,cm^{-3}$. At 3~Myr, the magenta and yellow
contours show densities of $4\times 10^3\,\rm H\,cm^{-3}$ and $2.5\,\rm
H\,cm^{-3}$, respectively.  The radiation pushes the gas from the outside and
disperses the outer region of the cloud. The dispersed gas moves at a speed of
$\sim 100\,\rm \, \rm km \, s^{-1}$ and is heated to a temperature of
$10^{4-5} \, \rm K$. } 
\label{fig:ZoomCloud}
\end{figure*}
\begin{figure*}
 \centering
 \leavevmode
 \includegraphics[width=0.93\textwidth]{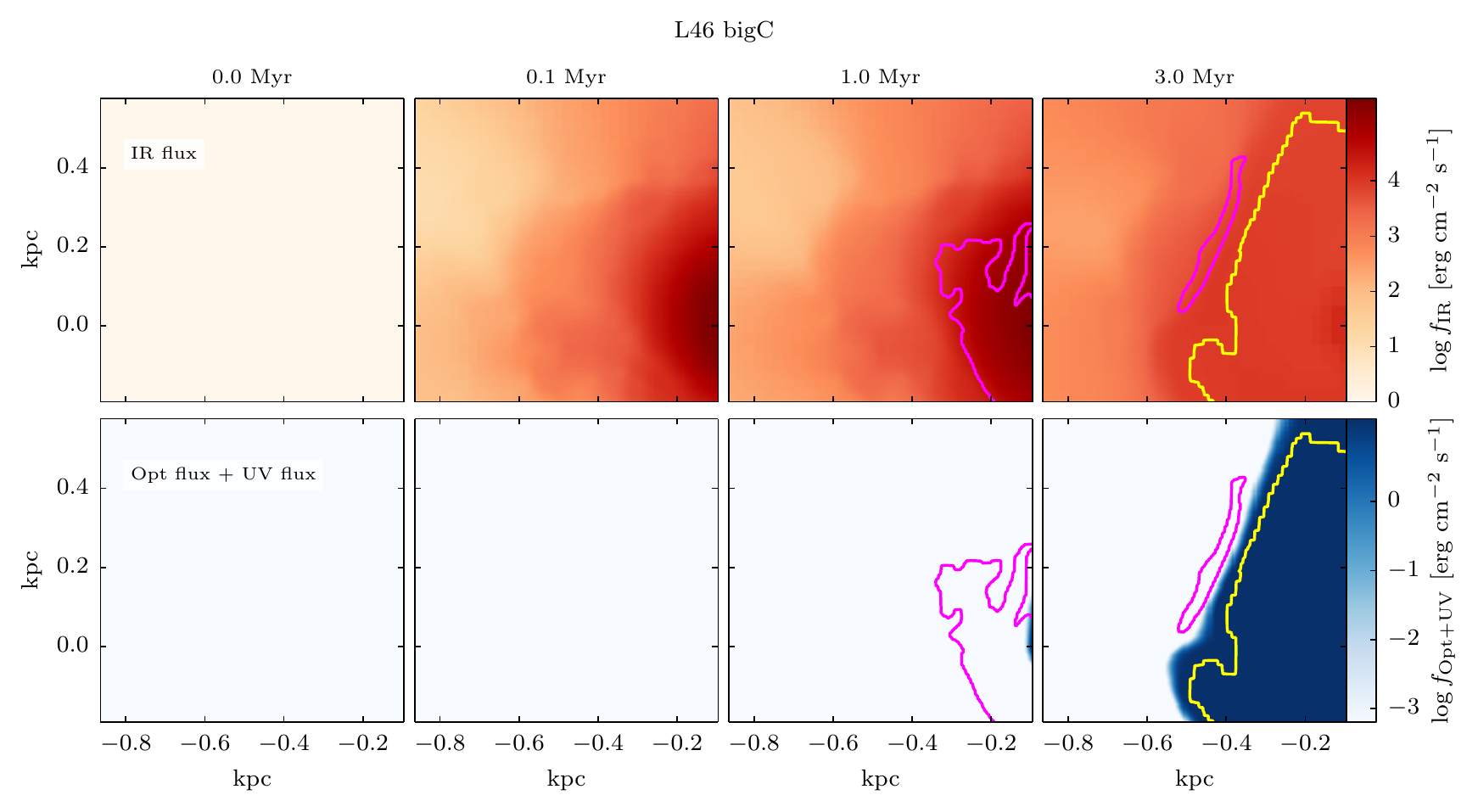}
 \caption{Slices of the IR (top row) and optical + UV (bottom row) radiation flux
 (over all directions) as a function of time for the same zoomed region as in
 Fig.~\ref{fig:ZoomCloud} for the same (\Le{46}\_\bigC) simulation. Here the UV
 flux is the sum of the flux from the individual photon groups UV1, UV2, and
 UV3. The IR flux is computed by considering both the streaming and trapped
 contributions to the IR energy density.  The same density contours as in
 Fig.~\ref{fig:ZoomCloud} are over-plotted to guide the eye.  While the IR
 radiation is completely penetrating the cloud after $\le$~0.1~Myr, the column
 density of the cloud is too high for the optical and UV radiation to penetrate
 the overdensity of the cloud.}
\label{fig:ZoomCloudPhotonGroups}
\end{figure*}
Fig.~\ref{fig:PropCloudLe46Max} illustrates the evolution of the mass-weighted
cloud mean density, temperature, and H\,{\sc ii} fraction for the different
\Le{46}\ simulations.  The lines show the mean values of all gas with
metallicity larger than $0.5$ Solar. This picks out the gas originally
belonging to dense clouds, since the gas within clouds is initialised with
Solar metallicity, while the diffuse gas outside them is initially metal-free.

Looking at the mean density evolution for the three simulations, we see that
the clouds start to disperse right from the beginning, causing the mean density
of the clouds to drop. The decrease in mean density is similar for the
different simulations.  With the mean density of the clouds declining with
time, the mean free path of the IR photons also increases and one therefore
expects the IR photons to scatter less within the gas.

For the \Le{46}\_\medC\ and \Le{46}\_\bigC\ simulations the mean temperature
rises within a very short amount of time ($\sim$~0.1~Myr) to $10^{4.3 -
4.5}$~K, whereas it takes $\sim$~10~Myr for the \Le{46}\_\smallC\ simulation to
reach similar values.  Additionally, the H\,{\sc ii} fraction of the clouds also
increases with time, with a roughly 30\% ionisation fraction for the gas within
the clouds at the end of the simulation. Photoionisation proceeds much
more slowly for the small cloud than for the more massive clouds. 

As seen in Fig.~\ref{fig:Mape46DensfixedFill} the early evolution of
the density evolution is sensitive to small-scale inhomogeneities. In
the \Le{46}\_\smallC\ simulation the photons manage to quickly destroy
the encompassing cloud and then escape through lower density channels
without efficiently heating, ionising, and pushing the higher density
gas.  For the bigger cloud simulations, on the other hand, the
radiation is trapped for longer within the encompassing cloud, leading to
the photons to interact with the higher density gas.  On long
timescales, the radiation has smoothed out the inhomogeneities, and
hence the evolution of the mean temperature and ionisation fraction is
similar for all the cloud masses, and converges to the same values
since the total masses are the same.

Fig.~\ref{fig:ZoomCloud} shows a close-up of a slice of the gas density, gas
temperature, HII fraction, and velocity from the \Le{46}\_\bigC\ simulation as
indicated by the red square in Fig.~\ref{fig:Mape46DensfixedFill}.  The
position of the quasar is just outside the zoomed-in slice, on the right-hand
side.  Fig.~\ref{fig:ZoomCloudPhotonGroups} displays the time evolution of the
radiation fluxes in two wavebands: IR and optical + UV, for the \Le{46}\_\bigC\
simulation in the same zoomed-in region as in Fig.~\ref{fig:ZoomCloud}.

At the start of the simulation (left panels of Fig.~\ref{fig:ZoomCloud} and
Fig.~\ref{fig:ZoomCloudPhotonGroups}), the gas is in a fractal two-phase
medium, with a cold, fully neutral, high-density component and a hot, fully
ionised low density component, with both components at rest.  

At $0.1\,\rm Myr$, the IR photons have already penetrated the whole cloud.
Where the IR photon flux is high, the dense gas has already started moving
outwards as indicated by the moderate velocities (10-50~$\rm km\,s^{-1}$) on
the right edge of the zoomed-in region.  During this early stage, the gas in
the high IR flux region is already at higher temperatures than the rest of the
gas within the cloud (second panel of the second row of
Fig.~\ref{fig:ZoomCloudPhotonGroups}) but has not yet shock-heated to very high
temperatures. Given the moderate temperatures well below dust destruction the
cloud is not yet transparent and the IR photons multi-scatter within it, giving
a boosted push from the inside-out.  At $0.1\,\rm Myr$, the cloud is still
fully neutral as the UV photons do not penetrate into the cloud (shown in
Fig.~\ref{fig:ZoomCloudPhotonGroups}) because the gas is too dense and
UV-shielded.

At $1\,\rm Myr$, the gas has begun to move away from the radiation source.  
To guide the eye, we have marked the outflowing gas at
$3\,\rm Myr$ with selected contours.  The magenta contours at density
$4000\,\rm H\,cm^{-3}$ mark the shock front.  At
$1\,\rm Myr$, the shock front is still neutral with a temperature of
roughly $10^3$~K and is traveling at a velocity of
$\sim 50\,\rm km\,s^{-1}$.  Given the mean density of the front at
$1\,\rm Myr$ of $\sim 4\times 10^{3}\,\rm H\,cm^{-3}$, the mean free
path, $1/(\kappa\,\rho)$, of IR photons is only $\sim 1.8\,\rm pc$,
which indicates the IR radiation is trapped and multi-scattering in
this region allowing the IR radiation to efficiently transfer momentum
onto the gas.  In the regions where the gas is dense and the IR flux
is high, the temperature of the gas is also higher compared to the
rest of the cloud indicating a shock.  Additionally, in the regions of
high IR flux, the IR radiation is mixing the multiphase gas and
creating a more uniform structure.
  
At the right edge of the cloud, i.e. in the direction of the source, the diffuse
gas is heated up to high temperatures ($\sim$~10$^6$~K), is fully ionised, and
has a velocity of $\sim 10^3 \, \rm km\,s^{-1}$.  When looking at the photon
flux of the different radiation groups (see
Fig.~\ref{fig:ZoomCloudPhotonGroups}), we see that, unlike the IR photons, the
UV and optical radiation still does not deeply penetrate the dense cloud.  We
will later see in more detail that the early evolution of dense clouds is
clearly dominated by the IR radiation, whereas the UV radiation plays a more
important role at later times.
% ---------------- Figures Velocity Evolution ----------------------- %
%
\begin{figure*}
\centering
\includegraphics[width=0.93\textwidth]{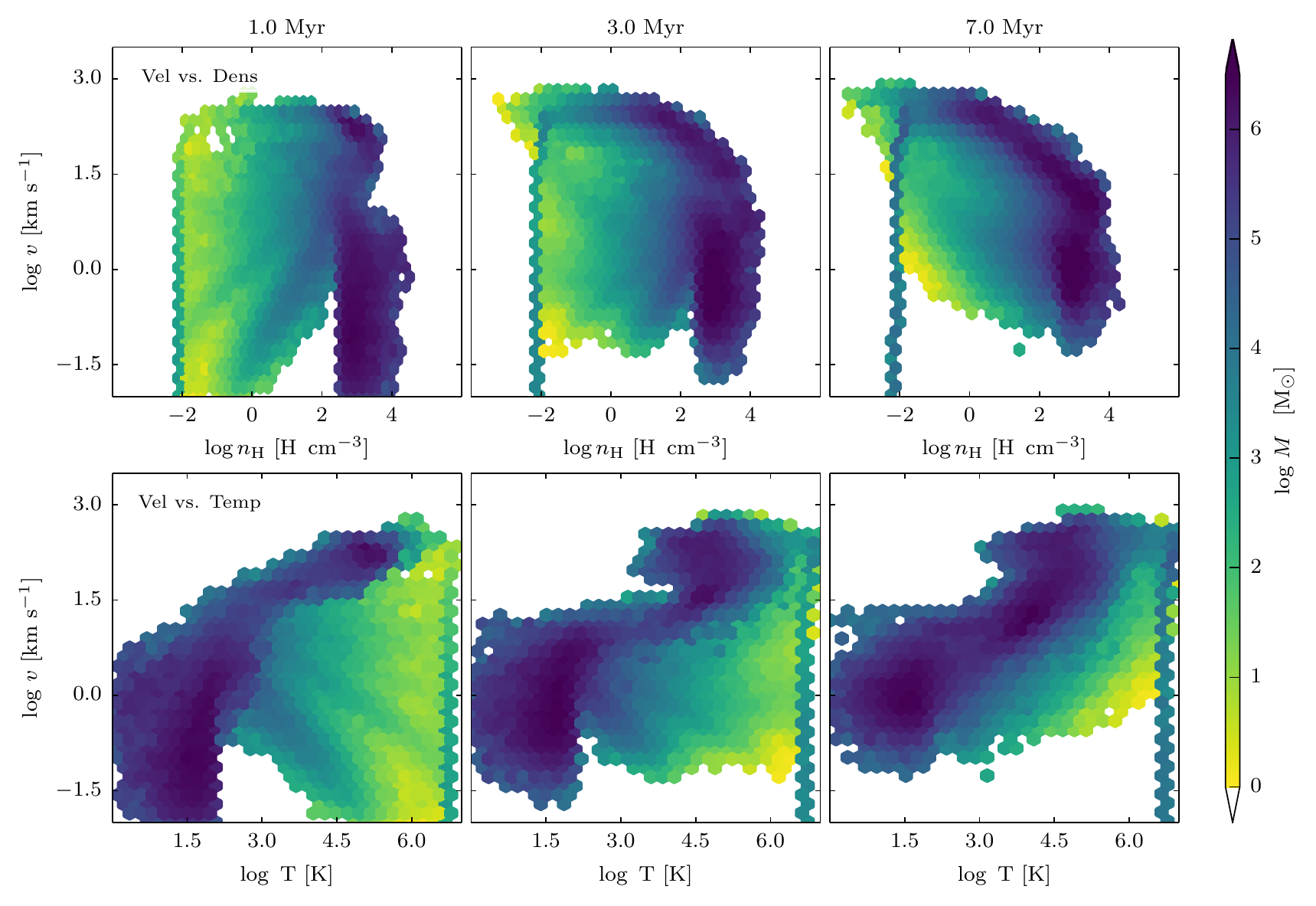}
\caption{\emph{Top:} Mass-weighted velocity versus density for the
\Le{46}\_\medC\ simulation. \emph{Bottom:} Mass-weighted velocity versus
temperature for the \Le{46}\_\medC\ simulation.  The points are coloured by the
total mass within each 2D-histogram cell.  The different columns show different
times as labeled.  The dense cold gas is accelerated by the radiation, expands and
results in a broad, diffuse, and fast wind.  The highest velocities for the
dense gas is lower at late times, since it corresponds to more distant clouds
receiving a lower photon flux. The fastest velocities are reached for gas with 
temperature below dust destruction.}
\label{fig:StuffVsVel}
\end{figure*}
%
% ---------------- Figures Velocity Evolution ----------------------- %

When the radiation illuminates dense clouds, the gas is dispersed starting from
the illuminated side, as seen in the density slice at $3\,\rm Myr$. To guide
the eye, we have again marked the outflowing gas with two selected contours. The
magenta contour at density $4\times 10^3\,\rm H\,cm^{-3}$ marks the shock
front, while the yellow contour at density $2.5\,\rm H\,cm^{-3}$ marks the
outflow tail before the density drops to the background density.  The dispersed
gas moves away from the source at a velocity of $\sim 50-150\,\rm km\,s^{-1}$
and has a temperature of around $\sim \,10^5 \,\rm K$.  The gas flowing away
from the dispersed cloud still has only a marginally smaller density than the
cloud, but is smoothed out by the IR radiation A large portion of the outflow
behind the density ridge (magenta contour) is fully ionised whereas the cold
cloud preceding the outflow is still neutral (see also
Fig.~\ref{fig:PropCloudLe46Max}).  As seen in
Fig.~\ref{fig:ZoomCloudPhotonGroups}, the UV radiation has only just reached
into the tail edge of the outflowing gas (yellow contour), started to ionise
the gas from the outside and heat it via photoionisation, and hence push it
further from the back end.  The rest of the gas is thus ionised via collisional
ionisation.  We will see in Section~\ref{subsec:velEvol} that gas is
accelerated by the radiation pressure from the UV photons and by
photoionisation heating, but UV photons contribute mostly to the overall
radiatively-driven wind once they are reprocessed in the IR.  The gas reaches
temperatures of around $\sim 10^5\,\rm K$ and velocities up to $500\,\rm
km\,s^{-1}$ beside the neutral gas, but still within the smooth outflowing gas
from the clouds (left from the yellow contour in Fig.~\ref{fig:ZoomCloud}).  In
the regions where the gas mixed with the background gas (right from the yellow
contour), the gas reaches temperatures up to $10^{6.5}$~K and velocities of up
to $1000\,\rm km\,s^{-1}$.  At $t=3\,\rm Myr$, the cloud size with a mean
density of $\sim 10^{3}\,\rm H\,cm^{-3}$ is now comparable to the mean free
path of the IR radiation, $\sim30\, \rm pc$, causing the IR flux, and thus the
influence of the IR photons, to decrease.

In summary, we find that for smaller, more fragmented clouds encompassing the
radiation source, the radiation has a tendency to escape, and hence it is less
efficient at heating and ionising the gas compared to larger and more coherent
surrounding structures which efficiently trap the radiation. Focusing on a
single cloud close to (but separated from) the source of radiation, we see that
the IR efficiently penetrates the cloud and pushes from the inside out,
smoothing out inhomogeneities, while the UV (and optical) radiation cannot
penetrate and acts more by pushing on and heating into the side of the cloud.

\subsubsection{Velocity Evolution}
\label{subsec:velEvol}
\begin{figure}
\centering
\includegraphics[width=0.93\columnwidth]{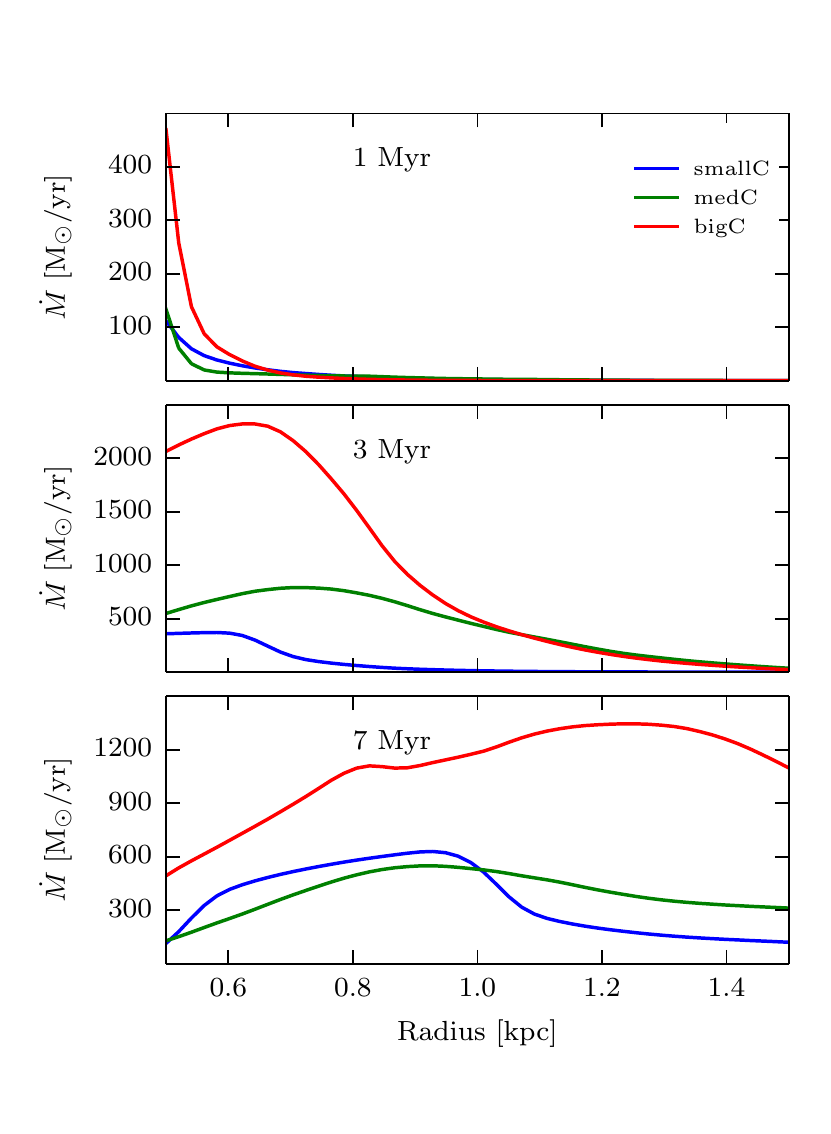}
\caption{Mass outflow rate as a function of radius for three different times 1,
3 and 7 Myr from top to bottom (same time as those used in
Fig.~\ref{fig:StuffVsVel}) for the \Le{46} simulations.  The radiation causes
the gas to move out of the galaxy, reaching mass outflow rates of up to
$500-1200\, \rm M_{\odot}\, yr^{-1}$ after 7~Myr depending on the cloud
encompassing the source.   The outflow rate is larger the bigger the
encompassing cloud because the radiation is  trapped for longer within the big
clouds and hence has longer time to impart momentum onto the gas.} 
\label{fig:MassFlow}
\end{figure}
Fig.~\ref{fig:StuffVsVel} shows mass-weighted velocity-density diagrams and
velocity-temperature diagrams for the \Le{46}\_\medC\, simulation at different
times (1~Myr, 3~Myr, and 7~Myr from left to right).  Three regions can be
highlighted in the velocity-density diagram.  At low densities, there is a
range of velocities in distinct intervals.  In light of
Fig.~\ref{fig:ZoomCloud}, we see that the low density gas far from the source
is not accelerated while that closer to the source is already accelerated close
to $10^3\, \rm km\, s^{-1}$.  At the highest densities, some of the gas has low
to moderate velocities up to $100 \, \rm km\, s^{-1}$.  Again, comparing to
Fig.~\ref{fig:ZoomCloud}, the almost zero velocity gas is on the far side of
the source and has not yet been accelerated.  And finally, in the mid density
range (between 6 and 150~H~cm$^{-3}$) diagonal stripes of mass of
$\sim$~10$^4$~M$_\odot$  arise due to the dispersion of the fast-moving high
density gas to lower densities.  Because of the dispersion and mixing with the
background at rest, the dispersed gas also slows down building the diagonal
stripes observed.  

The highest velocity gas (tip of the velocity-density diagram) shows
an anti-correlation with density.  This corresponds to gas near the
source, where the radiation from the source disperses the clouds to
lower densities, which reach the highest velocities.  Lower density
gas is moved earlier by the outflow created by the photon-gas
interaction whereas it takes longer for high-density gas to reach the
same velocities.  The high-density gas eventually also reaches high
velocities resulting in the whole cloud to move, which leads to the
destruction of the whole disc.  Comparing the velocity-density
diagrams at different times, we see that at earlier times, gas at high
densities has faster velocities than at later times, which is caused
by several factors.  First, the outflow front reaching an overdensity
must decelerate while interacting with the gas from the cloud.
Secondly, as we already have seen above, the high density regions are
dispersed, leading to lower densities and higher velocities as seen at
$7\, \rm Myr$.  Finally, the flux of photons decreases with distance
from the source, and late times correspond to clouds farther away from
the central source, which receive a smaller flux of photons.

The mass-weighted temperature versus velocity evolution shows that the
radiation accelerates the cold and dense gas which expands further into a more
diffuse and broader wind.  Then, an increasing amount of gas reaches higher
temperatures, resulting in a high mass fraction of high-speed hot gas at the
end of the simulation.  At later times, the dense and cold gas has lower
velocities than at earlier times, as it corresponds to more distant clouds
receiving a lower photon flux.  The hot  ($T>10^5\, \rm K$) gas is optically
thin because it is fully ionised and  the dust is destroyed in it.  Therefore,
the bulk of the high velocity gas ($v>100\,\rm km\, s^{-1}$) corresponds to
intermediate values of temperature of a few $10^4\,\rm K$.

We have measured the velocities of the gas for the two other simulations (not
shown here), and they have a very similar evolution.  The main difference is in
the intermediate density (1-100~H~cm$^{-3}$) and temperature (10$^3$-10$^5$~K)
range, where the gas velocity is higher with  increasing cloud size around the
source: $100-500 \, \rm km \, s^{-1}$, $100-600 \, \rm km \, s^{-1}$, and
$200-1000\,\rm km\, s^{-1}$ for the \smallC, \medC, and \bigC\ simulations,
respectively.

We now measure the mass outflow rate  with
\begin{equation} 
\dot{M}_{\rm gas} = \oiint \rho \,\vec{v} \cdot \hat{\vec{r}}\, \mathrm{d}S 
= \sum_{i \in \mathrm{shell}} m_{i}\, \vec{v}_{i} \cdot \hat{\vec{r}}_i
\,{S\over V} \, ,
\label{massflux}
\end{equation}
by considering only the outward flow (cells with $\vec{v}_i \cdot
\hat{\vec{r}}_i>0$) across a spherical shell of radius $r$, where $i$ denotes
the index of a cell within the spherical shell of surface $S$ and volume $V$.
Finally, $\hat{\vec{r}_i}$ is the unit vector of the cell with velocity
$\vec{v}_i$.  Here, we adopt a shell of thickness 0.25~kpc.
Fig.~\ref{fig:MassFlow} shows the mass outflow rate as a function of radius for
the \Le{46} simulations measured at different times. The radiation pressure
causes the gas to  move out, reaching outflow rates of up to 500 to
2400~M$_{\odot}\,\rm yr^{-1}$ depending on the initial cloud setup.  The mass
outflow rate is always larger the bigger the encompassing cloud around the
source, due to the radiation being trapped for longer within the large clouds.
The values of the mass outflow rates of $1000- 2000\, \rm M_{\odot}\, yr^{-1}$
measured in the largest cloud simulation at $t\ge3\,\rm Myr$ as well as the
high velocities of $\sim 1000\,\rm km\,s^{-1}$ are close to those measured by
\citet{Tombesi+2015}.  As already stated above, the collapse time of the cloud
encompassing the source is $\sim$~4~Myr, which hence dynamically influences the
mass outflow of the galaxy, at least towards the end of the simulation.  Thus,
our predictions of the mass outflow rates are optimistic and have to be tested
with simulations including gravity. We leave this to future work. 

\subsection{Effects of Different Photon Groups on the Cloud Evolution}
\label{subsec:EffectPhotonGroups}
\begin{figure*}
 \centering
 \leavevmode
 \includegraphics[width=0.93\textwidth]{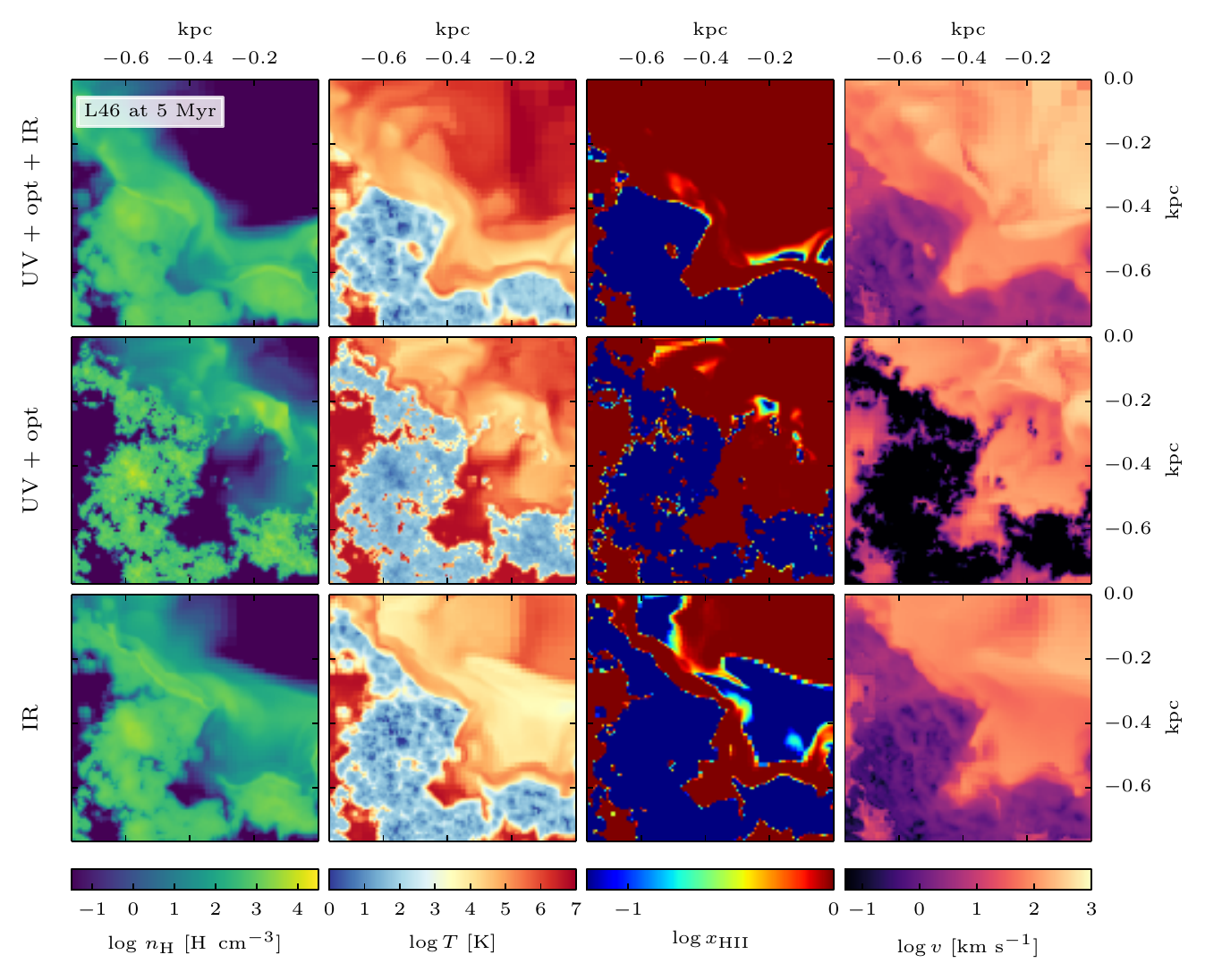}
 \caption{Zoomed-in slices of a cloud region in the \Le{46}\_\medC\ simulation
(red square of the middle row in Fig.~\ref{fig:Mape46DensfixedFill}) at 5~Myr,
showing, from left to right, maps of density, temperature, H\,{\sc ii}
fraction, and velocity.  The source is at the coordinate origin at the top
right corner each image.  The different rows show the same simulation with
different photon groups included, with, from top to bottom: all photon groups
(UV + Opt + IR), excluding the IR photons (UV + Opt), and finally including
only the IR photons (IR). The dispersion of the cloud is driven by a
complicated interplay between the IR and UV radiation. At early times, the
dominant contribution is however the IR radiation. } 
\label{fig:ZoomCloudIndividial}
\end{figure*}

To better determine the specific contribution of each photon group, we
have performed simulations of the same density distribution and quasar
luminosity for the medium cloud size simulation (\Le{46}\_\medC) where
we excluded certain photon groups. The density, temperature, H\,{\sc ii} 
fraction, and velocity maps of these simulations at $5\,\rm Myr$ can
be seen in Fig.~\ref{fig:ZoomCloudIndividial}.  The position of the
quasar source, at the coordinate origin, is at the top right corner of
the images.  In the top row, all the photon groups are included.  In
the middle row, the IR radiation is excluded, and in the bottom row
only the IR radiation is included (i.e.  the UV groups and optical are
excluded).

Comparing the different rows in Fig.~\ref{fig:ZoomCloudIndividial}, we
see that each photon group contributes to dispersing the dense gas,
but the IR contribution dominates as the outflow is clearly more
advanced in the IR-only run than in the run with only UV and optical.
However, even if the IR photons are most important in the gas
dispersion, the effect of the UV radiation is non-negligible.

Comparing the middle row with the bottom row of
Fig.~\ref{fig:ZoomCloudIndividial}, we see that the main difference between the
effects of IR and UV + optical photons is that the IR radiation plays the role of
smoothing out the dense gas, especially the regions which the UV and optical
radiation cannot reach.  When not including the IR photons, the cloud structure
is maintained with a similar multiphase state to that of the initial
conditions.  Generally, the UV photons only manage to disperse and ionise the
gas at the shock front and do not as efficiently mix and disperse the gas of
the cloud with that of the background, but instead push the over-dense gas by
direct radiation pressure and photoionisation.  The inability of UV photons in
efficiently mixing the multiphase gas is most apparent when looking at the
temperature (and velocity) structure. 
\begin{figure}
\centering
\includegraphics[width=\columnwidth]{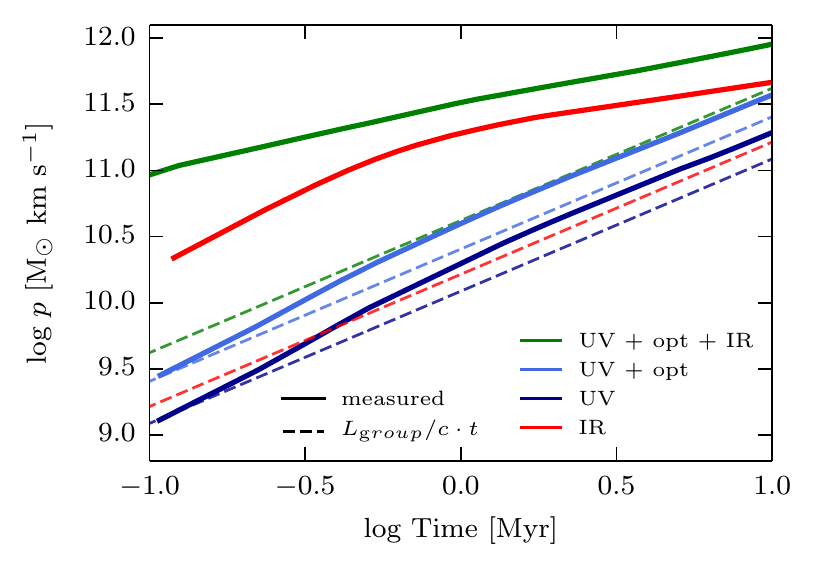}
\caption{Evolution of the total momentum for simulations where the contribution
of different photon groups are included: IR only (red lines), UV only (dark
blue lines), UV and optical (light blue lines), and all groups (green lines).
For the simulations, the exact same initial conditions as for the
\Le{46}\_\medC\ simulation, with all the photon groups included (solid green
line), are used.  The dashed lines show $(L_{\mathrm group}/c) \,t$, where $t$
is the time elapsed since the start of the simulations and $L_{\mathrm group}$
is the luminosity in the  photon group bands used in the simulation with the
same colour.  Through multiple scatterings on the dust, the IR radiation
imparts many times a momentum $L_{\mathrm group}/c$ onto the gas, thus greatly
boosting the total momentum transferred to the gas.  The main contribution to
the total momentum from the UV and optical photons comes from the reprocessed
UV photons into IR photons that then multi-scatter and impart a momentum boost
onto the gas.  Photoionisation heating has a small but non-negligible effect.
Finally, the optical photons give a small contribution to the momentum.}  
\label{fig:GroupMomentum}
\end{figure}

With only the IR radiation included, the gas of the cloud is much more mixed
with the background gas creating a more uniform density structure at the shock
front. This arises because the IR photons are isotropically pushing the gas
from the inside of the clouds and are, thus, responsible for the smoothing of the
multiphase density distribution and the puffing-up of the clouds.

Comparing the temperature slices of the rows, we observe that when the UV
photons are included, the temperature of the smoothed gas with densities of
$\sim 100 \,\rm H\,cm^{-3}$ and fully ionised is higher, due to the
photoionisation heating and extra momentum input.  We also see that the
increased velocity in the simulation incorporating all photon groups is due to
the combined contribution of all these photon groups.  As expected, the
ionization front is more advanced when both the UV (+ optical) and IR photons
are included compared to when only the IR radiation is included in the
simulation.

Looking at the velocity maps from the different simulations at 5~Myr,
we see that when the IR radiation is excluded, only the
ionised hot gas ($10^4 - 10^{6.5}\, \rm K$) moves with large
velocities ranging from $100$ to $1000\, \rm km\,s^{-1}$.
However, with the IR photons, the neutral gas is also moving with a
velocity of up to $100\,\rm km\,s^{-1}$ where the gas is warm ($10^4\,\rm K$),
and $\simeq 10\,\rm km\,s^{-1}$ where the gas is cold ($<10^2\,\rm K$).
The IR photons are, hence,
capable of moving the dense, neutral gas, which the UV and optical
radiation cannot reach due to the high optical depth of the cloud.  We
see that when all photon groups are included, the wind, driven from
the central region, collides with the external parts due to the UV
photon heating and contributes to driving the wind on large-scales.

Fig.~\ref{fig:GroupMomentum} shows the evolution of the total momentum for the
same simulations (\Le{46}\_\medC) as shown in
Fig.~\ref{fig:ZoomCloudIndividial}: including all photon groups (solid green
line), excluding contribution from IR photons (solid light blue line).
Additionally shown, is a simulation only including the UV photons (solid dark
blue line). The dashed lines show $L_{\mathrm group}/c t$, where $t$ is
measured as the time passed since the beginning of the simulation and
$L_{\mathrm group}$ is the luminosity in the used photon group bands used in
the corresponding simulation (indicated with the same color). The ratio between
the solid and dashed lines shows how much the corresponding photon groups boost
the amount of momentum transferred to the gas. Hence, the IR photons are
capable of strongly boosting the momentum transfer due to multi-scattering.
The UV photons indirectly transfer momentum to the gas via photoionisation
heating and with this additionally boost the momentum transfer, however not as
strongly as the IR photons.  Comparing the simulations that use the different
photon groups shows that the inclusion of the IR photons to the optical and UV
photons produces a greater momentum boost than in the simulation with only the
UV and optical groups, especially at early times. This shows that the main
effect of the UV and optical photons on the momentum comes from the
dust-absorbed photons that are reprocessed into IR radiation and then
additionally boost the momentum transfer onto the gas via multi-scattering. 

The small difference between the total momentum of the simulation including
only the UV photons and the simulation including the optical as well as UV
photons confirms that the optical photons have a non negligible impact on the
evolution of the gas, since it doubles its total momentum (the fraction of
energy in the optical band is the same than in the UV). Obviously, the
contribution of all the photon groups is required to achieve the full momentum
(solid green line).

\subsection{Efficiency of the Photon-Gas Coupling}
\label{subsec:EffMomTransfer}
In order to quantify the efficiency with which the radiation couples to the gas
and transfers momentum to it, we define the mechanical advantage as the ratio between
the momentum input rate $\dot{p}$ and the instantaneous momentum from the
radiation source given by $L/c$, where $L$ is the bolometric luminosity of the
source.  We calculate the instantaneous momentum injection to the gas as $\dot
p=(p_{N}-p_{N-1})/\Delta t_{N}$, with $p_{N}$ the total gas momentum at one
given snapshot of the simulation, and $\Delta t_{N}$ the time interval between
two snapshots. As shown in Fig.~\ref{fig:Spectrum}, $\sim$~80\% of the total
bolometric luminosity is covered by the photon groups used in the simulations
(recall that we do not cover the hard X-ray band).  A mechanical advantage
above unity occurs when the photons boost the momentum transfer between the
radiation and the gas. This can happen when, the IR photons are multiply
scattered, as well as by photoionisation heating and subsequent shock
formation.  The reason we used the bolometric momentum in this calculation is
because it is consistent with subgrid models of BH feedback in the literature
\citep[e.g.][]{DeBuhr+2011,Choi+2014,Barai+2014,Zubovas&Nayakshin2014,
Costa+2014,Hopkins+2016}.  

\begin{figure}
\centering
\includegraphics[width=\columnwidth]{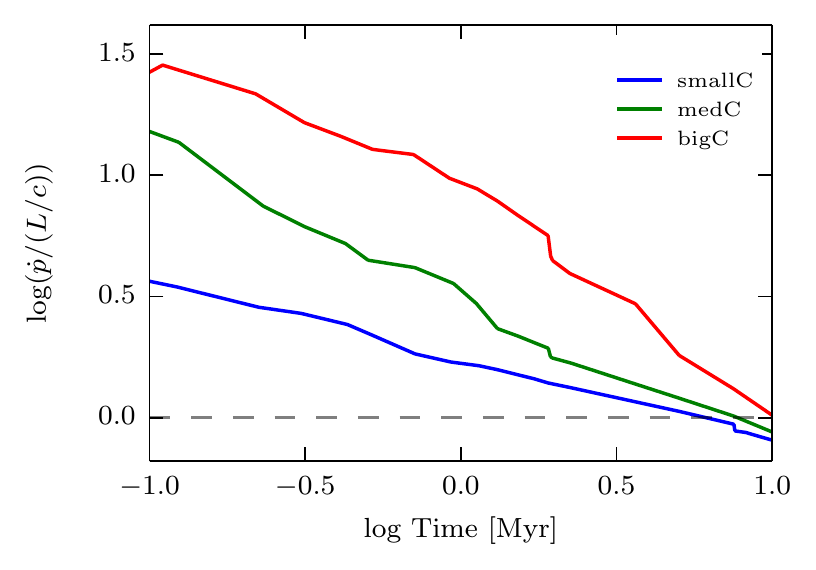}
\caption{Evolution of the mechanical advantage (momentum input rate
  $\dot p$ over $L/c$, where $L$ is bolometric luminosity) as a
  function of time for the \Le{46}\ simulations.  The mechanical
  advantage is above unity until $10$~Myr, it is larger the bigger the
  encompassing cloud, and it decreases with time.  For the \medC\ and
  \bigC\ simulations the mechanical advantage starts decreasing before
  the central cloud is fully destroyed as the efficiency of the
  momentum transfer is already less efficient once the radiation
  manages to carve a hole into the center of the galaxy.}
\label{fig:MechAdv46Max}
\end{figure}

Fig.~\ref{fig:MechAdv46Max} displays the evolution of the mechanical advantage
for the \Le{46}\_\smallC, \Le{46}\_\medC, and \Le{46}\_\bigC\ simulations.  For
all the simulations, the mechanical advantage decreases with time.  However,
the magnitude of the mechanical advantage is not the same for the three
different cloud size simulations, in particular in the early stages.  The
efficiency of the momentum transfer for the \Le{46}\_\smallC, \Le{46}\_\medC,
and \Le{46}\_\bigC\ simulations is bigger, the larger the cloud encompassing
the source of radiation. Because it takes more time to destroy larger clouds,
the photons are trapped and scatter for a longer time.  Indeed, a bigger
encompassing region around the quasar results in larger the momentum boost from
the radiation. For all the simulations, the radiation carves, right from the
beginning, a hole into the center of the galaxy, through which the radiation
can escape (Fig.~\ref{fig:Mape46DensfixedFill}).  This causes the mechanical
advantage to decrease with time as the photons injected by the source are more
likely to escape the system without scattering.

\begin{figure*}
\centering
\includegraphics[width=\columnwidth]{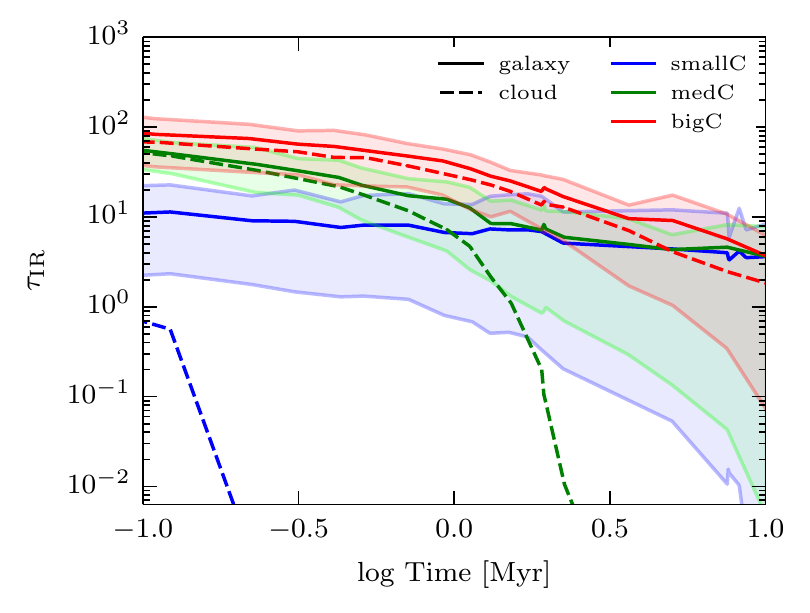}
\includegraphics[width=0.97\columnwidth]{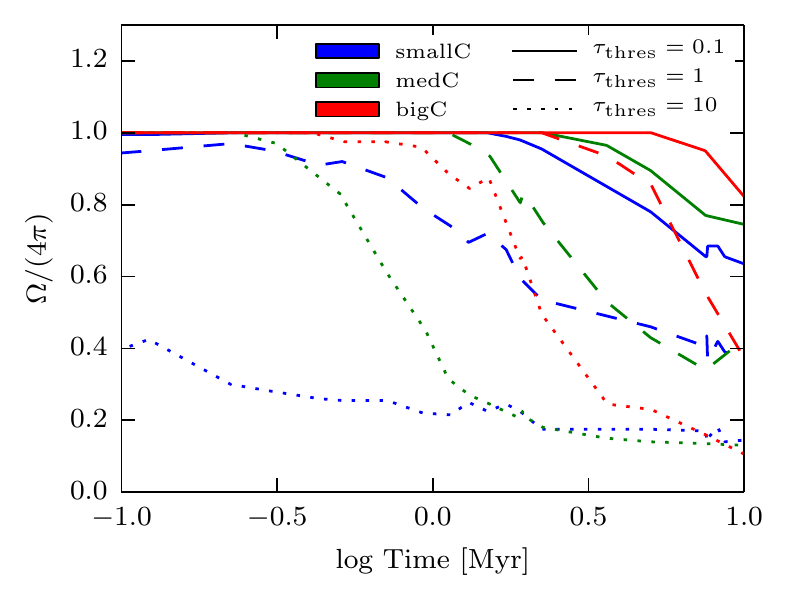}
\caption{\emph{Left:} Evolution of the optical depth $\tau_{\rm IR}$ as a
function of time for the \Le{46}\_\smallC, \Le{46}\_\medC, and \Le{46}\_\bigC\
simulations.  The lines show the mean values of the optical depth calculated by
sampling the sphere along 500 different lines of sight, over the galaxy disc
(up to 1.5 kpc, solid lines) or over the central cloud using a radius
corresponding to the largest fractal structure within the respective simulation
box (see Tab.~\ref{tab:sim}, dashed lines).  The shaded areas show the $\pm
1\,\sigma$ distribution of the optical depth over the galaxy.  The mean optical
depth of the three simulations depends on the cloud size and is larger the
bigger the clouds within the disc.  \emph{Right:} Fraction of solid angle over
the sphere covered by an optical depth greater than a threshold optical depth
$\tau_{\rm thres}$ of 0.1 (solid), 1 (dashed), or 10 (dotted). The optical
depth is calculated over the whole sphere. The figure shows that with
increasingly bigger clouds, the covering fraction of $\tau_{\rm IR}>1$ is
enhanced at given times and decreases at later times.}
\label{fig:PropLe46Max}
\end{figure*}
We see that the optical depth (or cloud size) plays a very important factor in
the evolution of the mechanical advantage.  In addition, dust destruction in
hot gas can play a role in suppressing the amount of momentum passed from
radiation to gas.  We have seen in Section~\ref{subsec:Clouds}, and especially
in Fig.~\ref{fig:PropCloudLe46Max}, that the gas is quickly heated to high
temperatures for the \Le{46}\_\medC\ and \Le{46}\_\bigC\ simulations whereas it
takes longer for the \Le{46}\_\smallC\ gas to reach similarly high
temperatures. It leads us to the conclusion that dust destruction has more
influence for the bigger cloud simulations compared to the \Le{46}\_\smallC\
simulation. However, as we will see in Section~\ref{subsec:OpticalDepth}, this
effect is not as important as the effect of the cloud size and gas density
surrounding the source. 

\subsection{Evolution of the Optical Depth}
\label{subsec:OpticalDepth}
For a better understanding of the mechanical advantage from the radiation and
the efficiency of photon-gas coupling, we measure the optical depth through the
disc for the quasar IR radiation as a function of time for the
\Le{46}\_\smallC, \Le{46}\_\medC, and \Le{46}\_\bigC\ simulations.  The IR
optical depth is defined as
\begin{equation}
\tau_{\rm IR} = \int \rho \; \kappa_{\rm IR}\; {\rm d} l = \sum _{i \in
  \rm{LOS}} \; \rho _i \; \kappa_{\rm IR}(T_i,Z_i) \; 
\Delta l_{\rm i}
\end{equation} 
where the opacity $\kappa_{\rm IR}$, function of temperature and metallicity,
is given in Eq.~(\ref{eq:opacity2}), $l$ is the line-of-sight (LOS) coordinate, and where
$\rho_i$, $T_i$ and $Z_i$ are respectively  the density, temperature and
metallicity of the cell of index $i$ along  the LOS, while $\Delta l_i$ is the
length of the LOS through the $i$th cell.

Fig.~\ref{fig:PropLe46Max} shows the evolution of the mean optical depth
calculated over 500 randomly selected LOS, uniformly sampling a sphere from the
centre of the disc, where the quasar source is located, up to a distance of
$1.5\, \rm kpc$.  

At the beginning of the simulation, the optical depth calculated over the
central cloud is $\sim 0.8$ times that of the optical depth calculated over the
whole disc for \Le{46}\_\medC\ and \Le{46}\_\bigC, while it is 0.1 that of the
disc for \Le{46}\_\smallC.  Once the radiation carves a hole into the central
cloud, the optical depth calculated over the cloud drops to zero.  This happens
later for bigger encompassing clouds.  Note that at $10\, \rm Myr$, the mean
optical depths from the source for the three simulations converge to the same
value of $\tau_{\rm IR}\simeq 4$.

As discussed above, there are two important phases for the IR radiation.
First, the radiation is trapped within an optically-thick region of the central
cloud and imparts momentum onto the gas, which dominates the evolution of the
outflow. Once the radiation has destroyed the central cloud, the IR radiation
is only trapped within the densest regions of the gas. We have seen that the
mean density of the cloud decreases with time, which, in turn, reduces the
optical depth of the gas and hence reduces the influence of the trapped photons
at later stages.  Therefore, in this second phase, the IR radiation plays a
less dominant role in pushing the gas out of the disc.  These two phases are
clear in Fig.~\ref{fig:PropLe46Max}: the cloud starts are high optical depth
with little decrease, and then there is a rapid drop.

The optical depths displayed in the left panel of Fig.~\ref{fig:PropLe46Max}
show considerable scatter, so that much radiation can escape even when the mean
optical depth is high.  The right panel of Fig.~\ref{fig:PropLe46Max}, the
fraction of the solid angle covered with $\tau_{\rm IR}$ larger than a
threshold optical depth $\tau_{\rm thres}$ is shown. The fraction of the solid
angle covered with $\tau_{\rm IR}>1$ gives an indication of the trapping of the
IR photons, i.e.  their multiple scattering, which in turns gives an indication
of the efficiency of photon to gas coupling.

The general evolution of the optical depth distribution for the
\Le{46}\_\smallC, \Le{46}\_\medC, and \Le{46}\_\bigC\ simulation helps to
understand the evolution of the mechanical advantage. At the start of the
simulations, before the central cloud is destroyed, the optical depth
distributions calculated over the cloud for the \Le{46}\_\medC and
\Le{46}\_\bigC\ simulations overlap, with mean optical depths of $\left\langle \tau _{\rm IR}\right\rangle
= 75 $ and 56 for the \Le{46}\_\bigC\ and \Le{46}\_\medC\ simulations,
respectively (left panel).  However, the \Le{46}\_\bigC\ simulation has more
scatter in the optical depth distribution compared to that of the
\Le{46}\_\medC\ simulation.  Additionally, the two simulations have the same
unity fraction of solid angle over the sphere covered by $\tau_{\rm IR}$ larger
than unity for the first $\sim$~1~Myr (right panel). The larger optical depth
for the \Le{46}\_\bigC\ simulation compared with the \Le{46}\_\medC\ simulation
leads to a larger mechanical advantage at the start of the simulation. The mean
optical depth of the two simulations decreases with time and they converge
after $\sim$~10~Myr.  The mean optical depth at the beginning of the
\Le{46}\_\smallC\ simulation, on the other hand, is much smaller ($\left\langle \tau _{\rm
IR} \right\rangle \sim$~11). Moreover, the fraction of solid angle around the quasar with
$\tau _{\rm IR} >1$ is at 80 \% from the start for \Le{46}\_\smallC. Thus, it
explains the small mechanical advantage for the small cloud simulation when
compared to the larger cloud simulations since a significant fraction (20 \%)
of all possible lines of sights are optically thin.

Indeed, the optical depth around the source is the important factor in
understanding how much momentum from the photons can be transferred to the gas.
However, as we have seen, channels of optically-thin gas can form within the
optically-thick layers of gas, reducing the expected amount of momentum that is
effectively transferred to the gas.  Following \citet{Hopkins+2011}, we write
the expected momentum boost as
\begin{equation}
\dot{p} = (1 + \eta \,\tau _{\rm IR}) \,\frac{L}{c} \quad ,
\label{eq:HopkinsSub}
\end{equation} 
where the factor of $\eta \,\tau _{\rm IR}\, L / c$ accounts for the momentum
boost passed onto the gas by the total number of IR scattering events.
Eq.~(\ref{eq:HopkinsSub}) includes a dimensionless ad hoc \textit{reduction
factor} $\eta$ that accounts for extra sources of momentum (e.g. UV
photo-heating, X-ray Compton scattering off the electrons, dust photo-electric
heating increasing $\eta$ above unity) and inhomogeneities in the gas
(decreasing $\eta$ below unity).  Hence, the reduction factor accounts for the
fraction of expected multiple scattering that effectively happen. Note that
\citet{Hopkins+2011} (and others) also use the mean optical depth when defining
$\eta$. 

\begin{figure}
\centering
\includegraphics[width=\columnwidth]{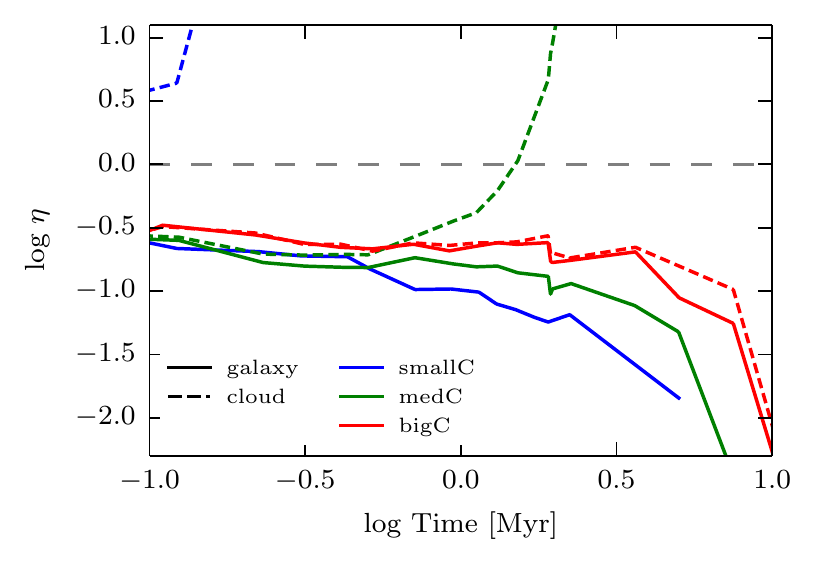}
\caption{Evolution of the reduction factor $\eta$, providing a measure of the
coupling efficiency of IR photons (see eq.~[\ref{eq:HopkinsSub}]) as a function
of time for the \Le{46} simulations.  The solid and dashed lines show the
reduction factor calculated with the mean optical depth within the galaxy
($\eta_{\rm gal}$) and within the central cloud ($\eta_{\rm cloud}$).
The values of the reduction factor become unphysical when $\tau_{\rm IR}$
falls below values around unity (cloud destruction).
}
\label{fig:eta}
\end{figure}

Given the change in momentum estimated from the simulation, we can determine
the reduction factor $\eta$ from Eq.~(\ref{eq:HopkinsSub}), and this is shown
in Fig.~\ref{fig:eta}.  The solid lines show the reduction factor calculated
with the mean optical depth within the galaxy ($\eta_{\rm gal}$), whereas the
dashed line show the reduction factor calculated using the mean optical depth
over the central cloud ($\eta_{\rm cloud}$).  

For all the three simulations, $\eta_{\rm gal}$ starts below unity starting at
$\sim$~0.2 for the \Le{46}\_\smallC\ and \Le{46}\_\medC\ simulation and
$\sim$~0.3 for the \Le{46}\_\bigC\ simulation and then decreases slowly with
time.  On the other hand, $\eta_{\rm cloud}$ starts with a similar value as
that of 
$\eta_{\rm galaxy}$ for the \Le{46}\_\medC\ and \Le{46}\_\bigC\ simulations,
whereas $\eta_{\rm cloud}$ for the \Le{46}\_\smallC\ simulation already starts
at a higher value due to the reduced optical depth over the cloud for this
simulation.  After $\sim$~1~Myr, $\eta_{\rm cloud}$ rises steeply in the
\Le{46}\_\medC\ simulation due to the optical depth dropping to zero around
that time.  The two reduction factors for the \Le{46}\_\bigC\ simulation evolve
very similarly, again due to the similar behaviour of the optical depth
calculated either over the whole galaxy or the encompassing cloud.

% ------------------ Figures Position -------------------------------------
\begin{figure*}
\includegraphics[width=0.95\textwidth]{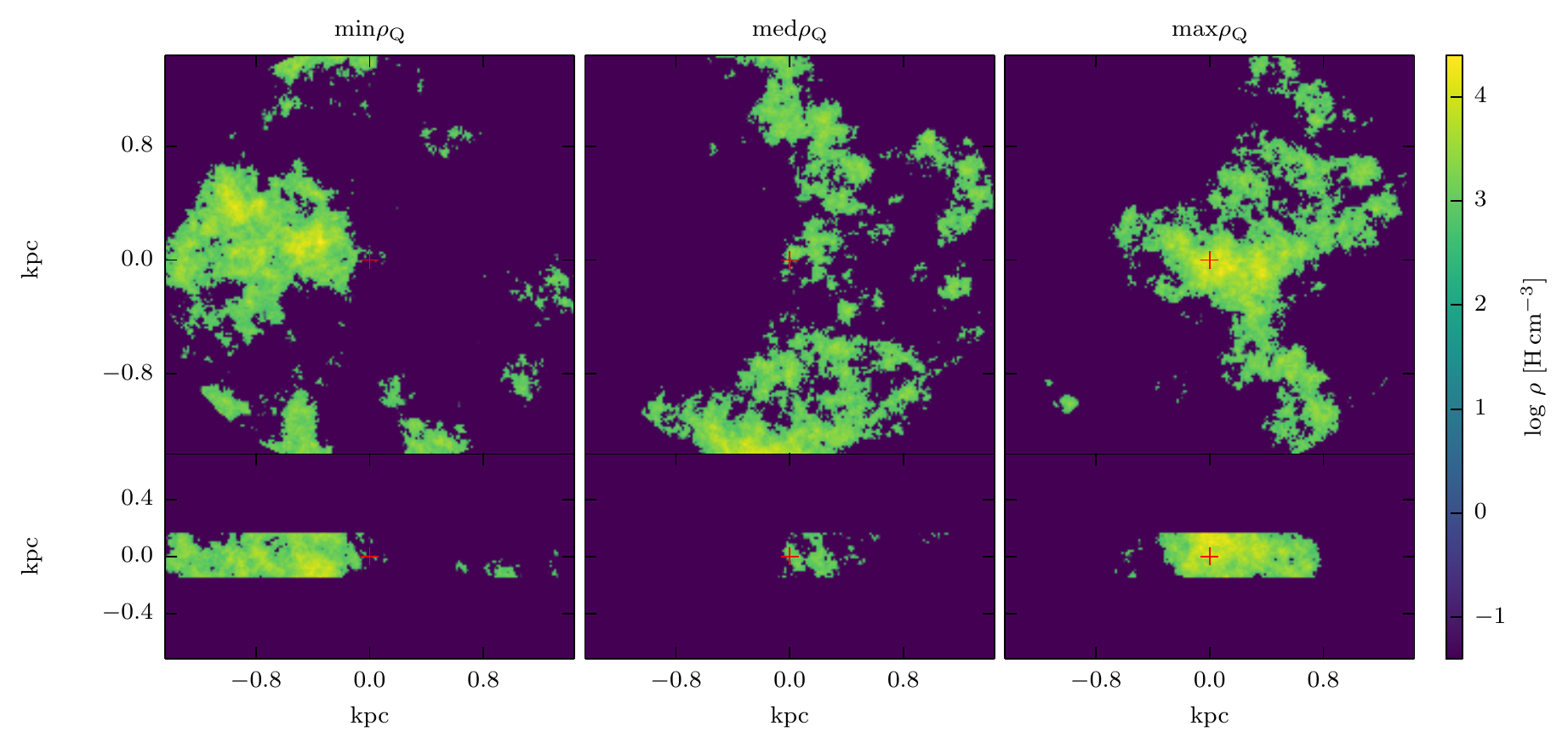}
\caption{Initial density distribution of the different environments for the
quasar with \Le{46}\_\bigC\_\textit{min$\rho_{\rm Q}$} (left panel),
\Le{46}\_\bigC\_\textit{med$\rho_{\rm Q}$} (middle panel),
\Le{46}\_\bigC\_\textit{max$\rho_{\rm Q}$} (right panel). The quasar position
is marked by a red cross at the center of each image.}
\label{fig:Le46PosIC}
\end{figure*}
% ------------------ End Figures Position ----------------------------------

It is important to keep in mind that Eq.~(\ref{eq:HopkinsSub}) does not hold
when the IR radiation is not trapped anymore and escapes without
multi-scattering i.e., when  $\left\langle\tau_{\rm IR}\right\rangle  < 1$,
which explains why the value of $\eta$ diverges in this regime.  This is mostly
the case for $\eta_{\rm cloud}$ when the central cloud is destroyed. Our
measurements show that, for the cases where $\left\langle\tau_{\rm
IR}\right\rangle \ge 1$, the measured mechanical advantage is well below unity
even when the source is placed in a region surrounded by a large optical depth.
It shows that the non-uniform structure of the ISM and the subsequent building
of low density channels as well as the building of a hole in the center of the
galaxy has a great influence in setting the reduction factor. 

This is in line with the argumentation by \citet{Krumholz&Matzner2009} and
\citet{KT2012, KT2013}, who use analytical arguments to show that the effective
optical depth can never be larger than a few, due to the tendency of photons to
escape through lower density channels, even though if the average optical depth
is much larger than unity.  Their findings have been challenged by
\citet{Davis+2014}, who showed that the results highly depend on the solver
used for the radiative transfer.  The flux-limited diffusion (FLD) method used
in \citet{KT2012} and \citet{KT2013} leads to a significantly lower efficiency
in accelerating the gas via radiation pressure than in the more accurate method
(variable Eddington tensor; VET) used by \citeauthor{Davis+2014} that showed a
stronger acceleration of the gas. \citet{RT2015} showed that the M1 method used
in our simulations lies somewhat between those of FLD and VET.  Using an
Implicit Monte Carlo radiation transfer scheme, \citet{Tsang+2015} find results
consistent with \citeauthor{Davis+2014}, demonstrating again the importance of
accurate radiative transfer in simulations of radiative feedback.
 
We have already seen that the initial stage, when the radiation is trapped
within the central cloud, plays a crucial role in the early acceleration of the
gas. We thus expect the IR radiation to be an important driver of the outflow
at early times.  However, the large IR luminosity rapidly destroys the cloud
encompassing the source, which leads to a decrease in the optical depth. This,
in turn, shortens the time the radiation is trapped within the central cloud,
where it can scatter sufficiently to impart a larger momentum boost onto the
gas.

\subsection{Effects of the Quasar Position}
\label{subsec:Position}

As radiatively-driven AGN winds are stronger when the source is embedded within
more massive clouds, we expect that the location of the source relative to the
cloud should also influence the momentum given to the gas.  We have
re-simulated the \Le{46}\_\bigC\ simulation, changing the underlying density at
the position of the quasar from a high-density environment (\Le{46}\_\bigC\
simulation shown before, hereby referred to as \textit{max$\rho_{\rm Q}$}
simulation) to a medium-density region (\textit{med$\rho_{\rm Q}$}) and to a
low-density environment (\textit{min$\rho_{\rm Q}$}). The regions are chosen
such that the average density, calculated over a region including the direct
neighbouring cells, is maximum ($\sim 8000$~$\rm{H} \, \rm{cm}^{-3}$), around
the mean ($\sim$~320~$\rm{H} \, \rm{cm}^{-3}$), and minimum ($\sim$~0.7~$\rm{H}
\, \rm{cm}^{-3}$), respectively.  The initial density distribution of the three
different simulations is shown in Fig.~\ref{fig:Le46PosIC}.

\begin{figure}
\includegraphics[width=0.99\columnwidth]{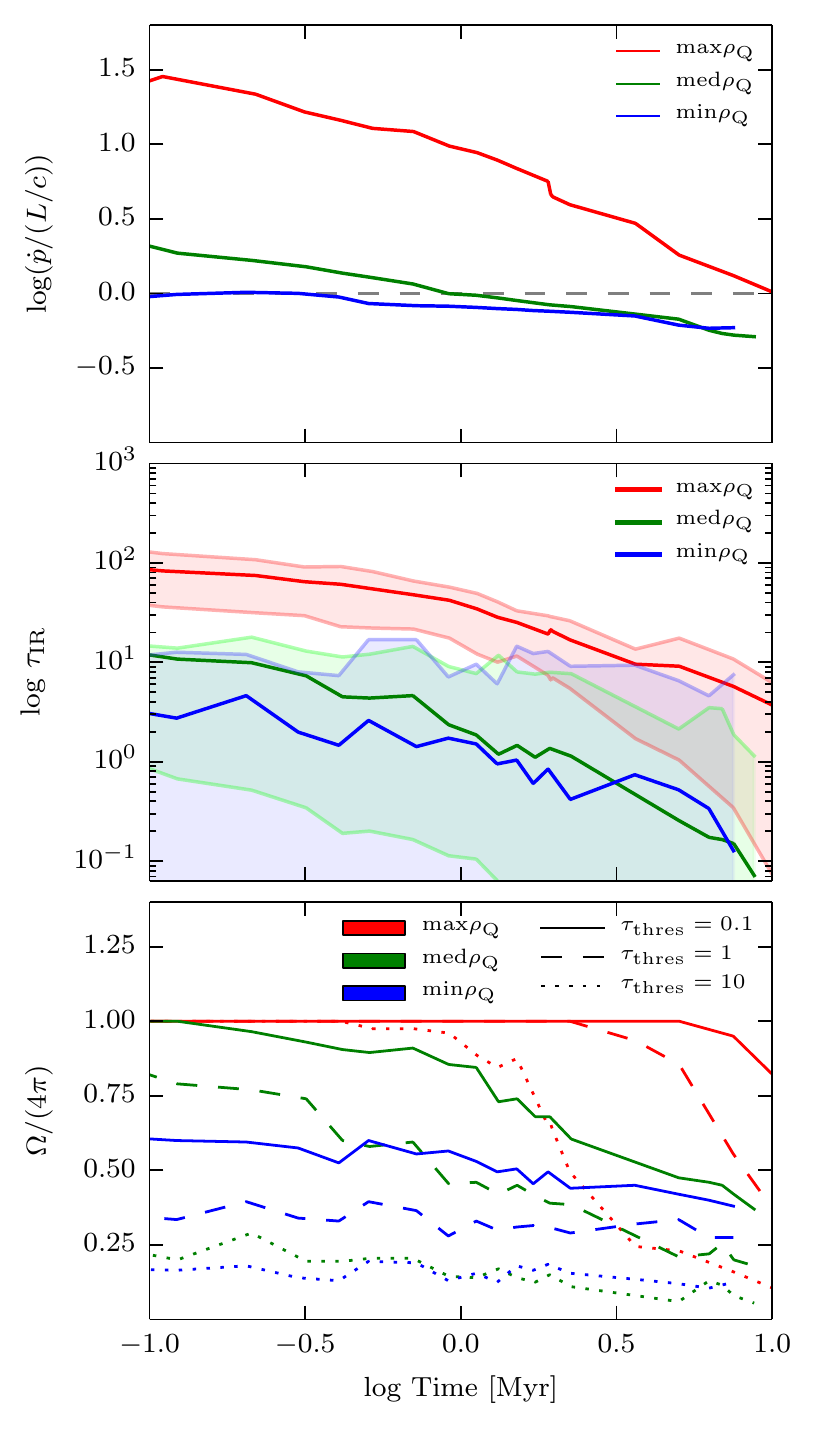}
\caption{\emph{Top}: Evolution of the mechanical advantage for the
\Le{46}\_\bigC\ simulations, where the position of the quasar within the
density field has been varied. The labels \textit{max}, \textit{mean}, and
\textit{min} stand for the quasar
position within the environment of  maximum, mean and minimum density,
respectively.
\emph{Middle}: Evolution
of the mean galaxy optical depth $\tau_{\rm IR}$ with the one $\pm \sigma$
standard deviation (shaded areas).  \emph{Bottom}: Fractions of solid angle
over the sphere covered with $\tau_{\rm IR}$ larger than a  threshold optical
depth $\tau _{\rm thres}$. We see that the
environment of the luminous source has a strong effect on the maximum boost
gained by the photons.} 
\label{fig:PropLe46Pos}
\end{figure}

The top panel of Fig.~\ref{fig:PropLe46Pos} displays the evolution of the
mechanical advantage for the three different quasar positions.  This figure
indicates that the maximum mechanical advantage reached in the simulation
increases by up to a factor 10 when the gas density at the position of the
quasar is increased from $\sim$~0.7~$\rm{H} \, \rm{cm}^{-3}$ to $\sim
8000$~$\rm{H} \, \rm{cm}^{-3}$.

The middle panel of Fig.~\ref{fig:PropLe46Pos} shows that the quasar position
has an important effect on the mean IR optical depth, which increases by a
factor of $\sim$~30 with increasing density at the quasar location.  According
to the lower panel of Fig.~\ref{fig:PropLe46Pos}, at the start of the
simulation, the fractions of solid angle covered with $\tau _{\rm IR} > 1$
increases from 40\% for the low gas density around the source to 100\% for the
run with the high gas density at the source location.

When the quasar is placed within the minimum density environment, the mean
optical depth is initially at $\left\langle\tau _{\rm IR}\right\rangle \sim 3$, but the solid angle
covered by an optical depth larger than $\tau _{\rm IR} = 1$ is only
$\sim$~30\%, and $\sim$~60\% of the IR photons have $\tau _{\rm IR} < 0.1$. A
significant fraction of the photons can hence free-stream out of the disc via
optically-thin channels and multi-scattering is negligible. The mechanical
advantage is at roughly unity for the first few $0.1\,\rm  Myr$ before
dropping off. Thus, there is no overall momentum boost in the beginning, but
since a fraction of the radiation escapes freely without interactions, another
fraction of the radiation must give a boosted momentum to the gas.

With the quasar placed at an intermediate density, $\sim$~80\% of the solid
angle around the source initially has $\tau_{\rm IR} > 1$, but a non-negligible
fraction of $\sim$~15\% of lines-of-sight have $\tau_{\rm IR} < 0.1$ and hence
are more or less free-streaming channels.  In this set-up, the mechanical
advantage is intermediate between the two other cases, starting at $\sim$~3,
steadily dropping, and going below unity after about $1\,\rm Myr$.
Thus, the probability of the photons scattering more than once is
much smaller when the local density around the quasar is intermediate compared
to when it has the largest value.  The environment of the source determines the
optical depth around the quasar, which influences the amount of momentum the
photons can impart onto the gas. If the source is already in an environment
where the covering fraction for $\tau_{\rm IR} > 1$ is small, the mechanical
advantage is much smaller than when the source is in an environment with a
large optical depth and the covering fraction for $\tau_{\rm IR} > 1$ is
higher.

\subsection{Comparison between Different Luminosities}
\label{subsec:Luminosity}
\begin{figure}
\includegraphics[width=0.99\columnwidth]{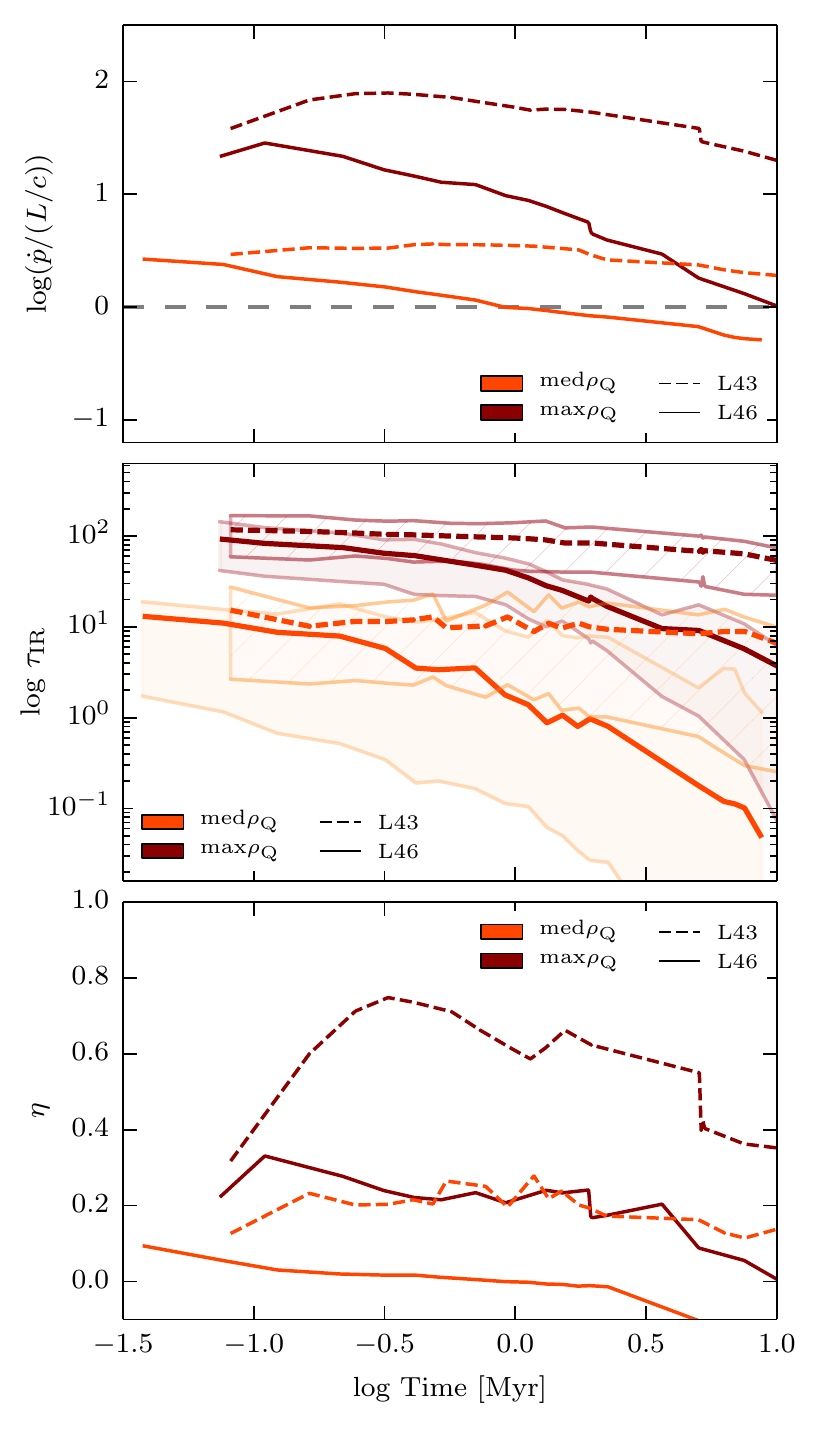}
\caption{\emph{Top}: Evolution of the mechanical advantage for low and high
quasar luminosities, respectively $L=10^{43}\, \rm erg\,s^{-1}$ (\Le{43}) and
$L=10^{46}\, \rm erg\,s^{-1}$ (\Le{46}), and two different local gas densities
(see Fig.~\ref{fig:Le46PosIC}).  \emph{Middle}: Evolution of the mean IR
optical depth with the $\pm1\,\sigma$ standard deviation (shaded areas for
\Le{46} simulations, hatched areas for \Le{43} simulations).  \emph{Bottom}:
Evolution of the reduction factor $\eta$. The mechanical advantage for the
lower luminosity simulation is higher than that of the higher luminosity
simulation with a larger value of $\eta$.  The higher luminosity leads to a
faster destruction of the central cloud compared to the lower luminosity,
leaving less time for the radiation to impart its momentum to the gas.}
\label{fig:PropDiffLum}
\end{figure}
We now study the effect of the quasar luminosity on the mechanical advantage.
We compare four simulations, using two different quasar luminosities,
$L=10^{43}$ and $10^{46}\, \rm erg\,s^{-1}$ (the \Le{43}\ and \Le{46}\
simulations, respectively), and two different quasar locations: intermediate
and high gas density around the source, for the largest cloud size simulation.

The top panel of Fig.~\ref{fig:PropDiffLum} shows the mechanical advantage of
the four simulations as a function of time. For a given density at the quasar
location, the mechanical advantage at early times (0.1 Myr) has similar values
for the two different luminosities.  However, in the higher luminosity case it
peaks and drops sooner than in the lower luminosity case.  

The reason for the higher mechanical advantage in the low luminosity
simulations is that the time it takes for the \Le{46}\ source to disperse the
encompassing cloud and to carve a hole into the centre of the disc is shorter
than for the \Le{43} source. This is clear when comparing the evolution of the
optical depth for low and high luminosity (middle panel of
Fig.~\ref{fig:PropDiffLum}).  The optical depth decreases faster with the
high-luminosity quasar than with the lower luminosity one.  Indeed, with the
lower luminosity quasar, the radiation has thus more time during which the
surrounding cloud is intact and the radiation multi-scatters inside it.  On the
other hand, with a more luminous source, the surrounding cloud is more quickly
dispersed, allowing the photons to escape, and giving a lower mechanical
advantage.

It is important to stress that the mechanical advantage is defined as the
fraction of the momentum change over the total bolometric momentum, where the
use of the bolometric momentum is consistent with subgrid models of black hole
feedback in the literature \citep[e.g.][]{Zubovas&Nayakshin2014,
Costa+2014,Hopkins+2016}. Hence, even though the mechanical advantage is higher
for the lower luminosity source, the total momentum at $\sim$~1~Myr for the
high luminosity source is $82\,\rm M_\odot\,km\,s^{-1}$ ($7.8\,\rm
M_\odot\,km\,s^{-1}$) and with this $\sim$~210 ($\sim$~360) times higher than
the total momentum in the low luminosity quasar simulation, with the
quasar placed in both simulations in a maximum (medium) density environment.

The reduction factor $\eta$ starts between $0.1$ and $0.3$ for the simulations
with the bright quasar and is roughly twice as  great in the simulations with
the faint quasar (bottom panel Fig.~\ref{fig:PropDiffLum}). The higher
reduction factor for lower luminosities is not only caused by the higher
mechanical advantage for the low luminosity simulation, but is also a
consequence of the different distributions of the optical depth.  The mean and
lower error bound of the optical depth distribution of the high quasar
luminosity simulation reaches much lower values than the corresponding values
in the low quasar luminosity run. Hence, while the high quasar luminosity run
produces a rapidly evolving outflow, with low-density channels that arise, in
the low luminosity run the gas has more time to mix with the surrounding gas,
leading to a smoother outflow, with no low-density channels arising.
Therefore, a larger fraction of photons manage to escape through lower density
channels for the higher luminosity simulation that then in turn lowers the
reduction factor.   

In summary, we find that the mechanical advantage is higher for lower
luminosities because the radiation manages to destroy the encompassing cloud
faster if the source is more luminous.  The reduction factor starts in a
similar range (0.1 - 0.3) for the two different luminosities, but then reaches
highest values for the simulation with low luminosity and quasar placed in a
maximum density environment. The reason for the higher reduction factor for
lower luminosities is that in the high luminosity run the radiation causes the
formation of low-density tunnels that are not formed in the lower luminosity
simulation due to the slower propagation of the wind. A larger amount of
low-density tunnels lead to a smaller efficiency of the momentum transfer
between the radiation and the gas and hence a smaller reduction factor for the
higher luminosity simulation.  However, the amount of momentum gained still
scales with the luminosity of the source.
%
% ------------------------------------------------------------------------------------------------------------------
\section{Discussion}\label{sec:discussion} % --------------------------------------------------------------------
While our simulations are well suited for studying the interactions between
radiation and gas, they come with caveats. We now discuss the shortcomings of
our simulations and possible implications and changes when taking them into
account.

First of all, including gravity in our simulations could have different,
possibly opposing, effects. The first effect is that gravity may slow down the
propagation of the outflows, as the gravitational potential will make it harder
for the gas to escape the galaxy, which in turn would decrease the mass outflow
rate out of the galaxy.  However, part of the outflow reaches velocities up to
$1000 \, \rm km \, s^{-1}$, which is fast enough for the gas to escape the
central encompassing cloud and even escape out of the massive halo. The escape
velocity out of the largest encompassing cloud of mean density of $\sim 100
\,\rm H\,cm^{-3}$ and radius $\sim 1.5\,\rm kpc$ is $\sim 450 \, \rm km \,
s^{-1}$.  The cloud collapse time, on the other hand, is $\sim$~9~Myr. Since
the outflow velocities of $1000 \, \rm km \, s^{-1}$ are greater than the
escape velocity out of the cloud and are reached on a faster timescale ($\le
5\,\rm Myr$) than the free-fall time of the central cloud, it is reasonable to
assume that the central cloud indeed gets destroyed and that the outflow
progresses further out of the galaxy.  

Within the disc, the influence of gravity is less clear.  In our simulation,
radiation is capable of carving a hole in the central regions of the galaxy. In
reality, part of the gas displaced by the radiation (that displaced
vertically above the now rotating disc) would likely fall back to
the centre, which would enhance the optical depth around the black hole and
thus could help to boost the momentum transfer from the radiation for a longer
time than observed in our simulations.

With the inclusion of radiative cooling, in addition to gravity, the collapse
of massive clouds would be enhanced, leading to a larger cloud mass of smaller
size.  This would likely strengthen the matter-radiation coupling due to the
larger optical depths of the clouds. However, radiative cooling (in combination
with gravity) could also lead to a fragmentation of the massive clouds (by
reducing the Jeans length) and thus speed up the formation of lower density
channels through which the radiation would escape.

In addition, our simulations explore the effect of a steady, uniform quasar,
located at the centre of the galaxy.  Actual black holes radiate with a
luminosity that is proportional to their accretion rate, and hence their
luminosities are not steady. Once the radiation manages to carve a hole in
the central regions of the galaxy, the luminosity of the quasar should drop to lower
values hence decreasing the amount of momentum transferred to the gas.

The interplay between the effect of gravity and the changing luminosity of the
quasar may lead to a self-regulating feedback cycle, where AGN activity pushes
the gas out of the central regions of the galaxy leading to starvation of the
black hole, shutdown of radiative emission and subsequent fall-back of the gas
onto the black hole.  Furthermore, black holes presumably accrete gas clouds
of varying masses and sizes.  Therefore, the amount of momentum (and feedback)
transferred to the galaxy, and the galactic mass outflow rate will vary with
time.

The inclusion of gravity, cooling, and the time variability of the quasar
luminosity, increases the complexity of the non-linear interplay between the
radiation and the gas, and we defer this study to future work.

Finally, it is still under debate whether an outflow driven by radiation from
the black hole may also lead to a (local) enhancement of star-formation due to
the compression of the clouds and the subsequent formation of more stars. We
have neglected the formation of stars and stellar feedback in our simulations.
Stars generally form in dense, cold gas regions. Removing gas mass from the
galaxy is thought to negatively impact the formation of stars. As deduced in observations
 \citep[e.g.][]{Cresci+2015}, it may also be plausible that the
compression of the gas during a burst of quasar activity also triggers star
formation.  Such a positive feedback effect has been shown for the
non-radiative modes of AGN feedback~\citep{Gaibler+2012, zubovasetal13,
bierietal15, bierietal16}.  We defer to future work a deeper discussion of the
possibility of triggered star formation due to quasar feedback.

\section{Conclusions}\label{sec:conclusion} % --------------------------------------------------------------------
Quasar-driven winds are powered by complex interactions between radiation and
gas. In most recent state-of-the-art hydrodynamical cosmological simulations
(but also in hydrodynamical simulations of isolated disc galaxies), quasar
feedback is approximated by depositing thermal energy within the resolution
element, where the efficiency of the radiation-gas coupling is represented by a
single parameter chosen to match global observations such as the SMBH
mass-bulge velocity dispersion~\citep{Ferrarese+Merritt2000}.  Additionally,
there is no consensus from these simulations on whether these AGN winds are
momentum-conserving or energy-conserving, although recent observations favour
energy-conserving winds from the nuclear accretion disc \citep{Tombesi+2015,
Fergulio+2015}.  Generally we refer to the outflow as energy-conserving if the
radiation is capable of efficiently boosting the momentum transfer between the
photons and the gas (i.e., $\dot p \gg L/c$), and to momentum-conserving if the
amount of momentum transferred to the gas is similar to the momentum flux of
photons provided by the source (i.e., $\dot p \approx L/c$).

Given the likely importance of AGN feedback in the evolution of massive
galaxies and the increasing amount of radio galaxy observations --- e.g., from
the Low-Frequency Array (LOFAR), the Australian Square Kilometer Array
Pathfinder (ASKAP), as well the Atacama Large Millimeter Array (ALMA), which
promise to provide us a better view of the ISM properties in these galaxies ---
it is time to improve our theoretical understanding of the mechanisms that
drive the momentum and energy transfer from the photons to the ISM in order to
properly quantify the role of AGN feedback in the evolution of galaxies.

We have performed idealised galactic disc RHD simulations to study the coupling
of photons with the highly multiphase galactic gas using different cloud sizes.
In the simulations, the emission, absorption, and propagation of photons and
their interaction with the gas via photoionisation, momentum transfer, and
absorption/scattering on dust, is followed self-consistently.

We find that radiation from a bright quasar is capable of driving a powerful
wind, with multi-scattering IR photons playing the dominant role, especially at
early times. The UV and optical photons never manage to reach the high density
regions of clouds due to the large optical depths. Therefore, the UV and
optical photons exert their direct push to the clouds from outside.  Some
of the UV and optical radiation is dust-absorbed and reprocessed into IR
radiation that can propagate much deeper into the dense gas. This reprocessed
IR radiation is the most significant contribution of the higher-energy
radiation to the total momentum of the gas (see Fig. \ref{fig:GroupMomentum}).

The mass outflow rates as well as the velocity reached by the galactic wind
depend on the structure of the ISM (and luminosity of the source).  Larger
clouds have greater IR optical depths,  hence longer time during
which IR photons are trapped within and scatter multiple times. This first phase plays a
crucial role in the early acceleration of the gas.  Once the radiation has
destroyed the central encompassing cloud, the IR radiation can stream out of
the galaxy through low density regions, decreasing the efficiency of the
momentum transfer from the photons to the gas.

Both the outflow rates of $\sim 500$ to 1000~M$_{\odot}\, \rm{yr}^{-1}$ and
high velocities of $\sim 1000\, \rm km\, s^{-1}$ that we measured in our
simulations with a large encompassing cloud around the quasar are close to
those observed by \cite{Tombesi+2015} and \cite{Fergulio+2015}. This agreement
favours winds that appear as energy-driven, which in our case are actually
radiatively-driven.  But as  mentioned above, our measured outflow rates are
optimistic predictions as we did not include gravity in our simulations.

The mechanical advantage, defined as the ratio of the imparted momentum rate
$\dot p$ and $L/c$, decreases with time, from $\sim3$ to 30 at the beginning of
the simulation to unity after $\sim$~10~Myr, thanks to the decreasing efficiency
of the photon-gas coupling. This mechanical advantage largely depends on the
size of the encompassing cloud as well as the position of the quasar (and the
quasar luminosity).  It varies by a factor of 10, depending on the size of the
encompassing cloud (50~pc to $1.5$ kpc) used in our simulations, with the IR
optical depths ranging from $\tau_{\rm IR} = $~10 to 100.  The position of the
quasar plays an important role in setting the amount of momentum transferred
from the photons to the gas, where the mechanical advantage changes by a factor
of 50, depending on whether the source is buried within the cloud or
illuminating it from the outside.  Thus, the IR optical depth $\tau_{\rm IR}$
(i.e., the cloud size and mass) around the quasar is the most important
criterion in determining the momentum injection rate at a given quasar
luminosity.

The reduction factor $\eta = [\dot p/(L/c)-1]/\tau_{\rm IR}$ is an empirical
estimate that accounts for extra sources of momentum (e.g., UV photoionisation
heating, X-ray Compton scattering off the electrons, dust photo-electric
heating, etc.) and inhomogeneities in the gas.  The measured reduction factor
never reaches values of unity, showing that the mechanical advantage never
reaches values as high as $\tau_{\rm IR}$.  The non-uniform structure of the
ISM and the formation of low density channels is responsible in setting this
low value of $\eta= 0.2-0.3$, that decreases significantly once the most
central cloud starts breaking up and becomes optically thin. Thus, the number
of scattering events of IR photons is roughly one quarter of the IR optical
depth.

The radiation emitted by the central quasar destroys the cloud encompassing the
source, which leads to a rapid decrease in the optical depth and reduction
factor.  We demonstrated that the destruction time is shorter for larger
luminosities, which leads to a smaller mechanical advantage. Additionally, the
fast evolution of the wind in the high luminosity simulation leads to the
formation of low density channels leading to a smaller efficiency of the
radiation-gas coupling. For the low luminosity simulation, on the other hand,
the evolution is slower which leaves the gas more time with to mix with the
surrounding gas and thus leads to a smaller amount of optically-thin gas.  This
in turn leads to a higher reduction factor for the lower luminosity simulation.
Nonetheless, large luminosities are required to obtain radiatively-driven
quasar winds with fast velocities and high mass outflow rates. 

In the future, this study will be extended to AGN radiation within
self-regulated turbulent discs including gravity, gas cooling, star formation,
stellar feedback and a model for dust creation and destruction.  

%==============================================
% Acknowledgments
%==============================================
\section*{Acknowledgments}
It is our pleasure to thank Romain Teyssier, Marta Volonteri, Pierre Guillard,
and Jonathan Coles for valuable discussions.  This work was granted access to
the HPC resources of CINES under the allocations c2015047421 made by GENCI.  RB
has been supported by the Institute Lagrange de Paris.  JR was funded by the
European Research Council under the European Union's Seventh Framework
Programme (FP7/2007-2013) / ERC Grant agreement 278594-GasAroundGalaxies, and
the Marie Curie Training Network CosmoComp (PITN-GA-2009-238356).  JS
acknowledges support from project 267117 (DARK) hosted by UPMC -- Sorbonne
Universit\'es, from JHU by National Science Foundation grant OIA-1124403.  This
work has made use of the Horizon cluster, hosted by the Institut
d'Astrophysique de Paris.  We warmly thank S.~Rouberol for running it smoothly.  

\vspace{-0.5cm}

\bibliographystyle{mn2e}
\bibliography{refs}

\begin{thebibliography}{90}
\expandafter\ifx\csname natexlab\endcsname\relax\def\natexlab#1{#1}\fi

\bibitem[{{Agertz} {et~al}\mbox{.}(2013){Agertz}, {Kravtsov}, {Leitner}, \&
  {Gnedin}}]{Agertz+2013}
{Agertz} O., {Kravtsov} A.~V., {Leitner} S.~N., {Gnedin} N.~Y., 2013, \apj,
  770, 25

\bibitem[{{Antonuccio-Delogu} \& {Silk}(2010)}]{Antonuccio-Delogu+Silk2010}
{Antonuccio-Delogu} V., {Silk} J., 2010, \mnras, 405, 1303

\bibitem[{{Barai} {et~al}\mbox{.}(2014){Barai}, {Viel}, {Murante}, {Gaspari},
  \& {Borgani}}]{Barai+2014}
{Barai} P., {Viel} M., {Murante} G., {Gaspari} M., {Borgani} S., 2014, \mnras,
  437, 1456

\bibitem[{{Bicknell} {et~al}\mbox{.}(2000){Bicknell}, {Sutherland}, {van
  Breugel}, {Dopita}, {Dey}, \& {Miley}}]{Bicknell+Sutherland+2000}
{Bicknell} G.~V., {Sutherland} R.~S., {van Breugel} W.~J.~M., {Dopita} M.~A.,
  {Dey} A., {Miley} G.~K., 2000, \apj, 540, 678

\bibitem[{{Bieri} {et~al}\mbox{.}(2015){Bieri}, {Dubois}, {Silk}, \&
  {Mamon}}]{bierietal15}
{Bieri} R., {Dubois} Y., {Silk} J., {Mamon} G.~A., 2015, \apjl, 812, L36

\bibitem[{{Bieri} {et~al}\mbox{.}(2016){Bieri}, {Dubois}, {Silk}, {Mamon}, \&
  {Gaibler}}]{bierietal16}
{Bieri} R., {Dubois} Y., {Silk} J., {Mamon} G.~A., {Gaibler} V., 2016, \mnras,
  455, 4166

\bibitem[{{Booth} \& {Schaye}(2009)}]{BoothSchaye2009}
{Booth} C.~M., {Schaye} J., 2009, \mnras, 398, 53

\bibitem[{{Bower} {et~al}\mbox{.}(2006){Bower}, {Benson}, {Malbon}, {Helly},
  {Frenk}, {Baugh}, {Cole}, \& {Lacey}}]{Bower+2006}
{Bower} R.~G., {Benson} A.~J., {Malbon} R., {Helly} J.~C., {Frenk} C.~S.,
  {Baugh} C.~M., {Cole} S., {Lacey} C.~G., 2006, \mnras, 370, 645

\bibitem[{{Capelo} {et~al}\mbox{.}(2010){Capelo}, {Natarajan}, \&
  {Coppi}}]{Capelo+2010}
{Capelo} P.~R., {Natarajan} P., {Coppi} P.~S., 2010, \mnras, 407, 1148

\bibitem[{{Choi} {et~al}\mbox{.}(2014){Choi}, {Naab}, {Ostriker}, {Johansson},
  \& {Moster}}]{Choi+2014}
{Choi} E., {Naab} T., {Ostriker} J.~P., {Johansson} P.~H., {Moster} B.~P.,
  2014, \mnras, 442, 440

\bibitem[{{Choi} {et~al}\mbox{.}(2012){Choi}, {Ostriker}, {Naab}, \&
  {Johansson}}]{Choi+2012}
{Choi} E., {Ostriker} J.~P., {Naab} T., {Johansson} P.~H., 2012, \apj, 754, 125

\bibitem[{{Cicone} {et~al}\mbox{.}(2014){Cicone}, {Maiolino}, {Sturm},
  {Graci{\'a}-Carpio}, {Feruglio}, {Neri}, {Aalto}, {Davies},
  {Gonz{\'a}lez-Alfonso}, {Hailey-Dunsheath}, {Piconcelli}, \&
  {Veilleux}}]{Cicone+2014}
{Cicone} C. {et~al.}, 2014, \aap, 562, A21

\bibitem[{{Ciotti} \& {Ostriker}(2007)}]{Ciotti+Ostriker2007}
{Ciotti} L., {Ostriker} J.~P., 2007, \apj, 665, 1038

\bibitem[{{Ciotti} \& {Ostriker}(2012)}]{Ciotti+Ostriker2012}
{Ciotti} L., {Ostriker} J.~P., 2012, in Astrophysics and Space Science Library,
  Vol. 378, Astrophysics and Space Science Library, {Kim} D.-W., {Pellegrini}
  S., eds., p.~83

\bibitem[{{Costa} {et~al}\mbox{.}(2014){Costa}, {Sijacki}, \&
  {Haehnelt}}]{Costa+2014}
{Costa} T., {Sijacki} D., {Haehnelt} M.~G., 2014, \mnras, 444, 2355

\bibitem[{{Cresci} {et~al}\mbox{.}(2015){Cresci}, {Marconi}, {Zibetti},
  {Risaliti}, {Carniani}, {Mannucci}, {Gallazzi}, {Maiolino}, {Balmaverde},
  {Brusa}, {Capetti}, {Cicone}, {Feruglio}, {Bland-Hawthorn}, {Nagao}, {Oliva},
  {Salvato}, {Sani}, {Tozzi}, {Urrutia}, \& {Venturi}}]{Cresci+2015}
{Cresci} G. {et~al.}, 2015, \aap, 582, A63

\bibitem[{{Croton} {et~al}\mbox{.}(2006){Croton}, {Springel}, {White}, {De
  Lucia}, {Frenk}, {Gao}, {Jenkins}, {Kauffmann}, {Navarro}, \&
  {Yoshida}}]{Croton+2006}
{Croton} D.~J. {et~al.}, 2006, \mnras, 365, 11

\bibitem[{{Daddi} {et~al}\mbox{.}(2010){Daddi}, {Bournaud}, {Walter},
  {Dannerbauer}, {Carilli}, {Dickinson}, {Elbaz}, {Morrison}, {Riechers},
  {Onodera}, {Salmi}, {Krips}, \& {Stern}}]{Daddi+2010}
{Daddi} E. {et~al.}, 2010, \apj, 713, 686

\bibitem[{{Davis} {et~al}\mbox{.}(2014){Davis}, {Jiang}, {Stone}, \&
  {Murray}}]{Davis+2014}
{Davis} S.~W., {Jiang} Y.-F., {Stone} J.~M., {Murray} N., 2014, \apj, 796, 107

\bibitem[{{Debuhr} {et~al}\mbox{.}(2011){Debuhr}, {Quataert}, \&
  {Ma}}]{DeBuhr+2011}
{Debuhr} J., {Quataert} E., {Ma} C.-P., 2011, \mnras, 412, 1341

\bibitem[{{Debuhr} {et~al}\mbox{.}(2010){Debuhr}, {Quataert}, {Ma}, \&
  {Hopkins}}]{DeBuhr+2010}
{Debuhr} J., {Quataert} E., {Ma} C.-P., {Hopkins} P., 2010, \mnras, 406, L55

\bibitem[{{Di Matteo} {et~al}\mbox{.}(2008){Di Matteo}, {Colberg}, {Springel},
  {Hernquist}, \& {Sijacki}}]{DiMatteo+2008}
{Di Matteo} T., {Colberg} J., {Springel} V., {Hernquist} L., {Sijacki} D.,
  2008, \apj, 676, 33

\bibitem[{{Di Matteo} {et~al}\mbox{.}(2005){Di Matteo}, {Springel}, \&
  {Hernquist}}]{DiMatteo+Springel+2005}
{Di Matteo} T., {Springel} V., {Hernquist} L., 2005, \nat, 433, 604

\bibitem[{{Draine} \& {Salpeter}(1979)}]{Draine+Salpeter1979a}
{Draine} B.~T., {Salpeter} E.~E., 1979, \apj, 231, 77

\bibitem[{{Dubois} {et~al}\mbox{.}(2012){Dubois}, {Devriendt}, {Slyz}, \&
  {Teyssier}}]{duboisetal2012}
{Dubois} Y., {Devriendt} J., {Slyz} A., {Teyssier} R., 2012, \mnras, 420, 2662

\bibitem[{{Dubois} {et~al}\mbox{.}(2014){Dubois}, {Pichon}, {Welker}, {Le
  Borgne}, {Devriendt}, {Laigle}, {Codis}, {Pogosyan}, {Arnouts}, {Benabed},
  {Bertin}, {Blaizot}, {Bouchet}, {Cardoso}, {Colombi}, {de Lapparent},
  {Desjacques}, {Gavazzi}, {Kassin}, {Kimm}, {McCracken}, {Milliard},
  {Peirani}, {Prunet}, {Rouberol}, {Silk}, {Slyz}, {Sousbie}, {Teyssier},
  {Tresse}, {Treyer}, {Vibert}, \& {Volonteri}}]{HAGN}
{Dubois} Y. {et~al.}, 2014, \mnras, 444, 1453

\bibitem[{{Faucher-Gigu{\`e}re} \& {Quataert}(2012)}]{FGQuasar2012}
{Faucher-Gigu{\`e}re} C.-A., {Quataert} E., 2012, \mnras, 425, 605

\bibitem[{{Ferrarese} \& {Merritt}(2000)}]{Ferrarese+Merritt2000}
{Ferrarese} L., {Merritt} D., 2000, \apjl, 539, L9

\bibitem[{{Feruglio} {et~al}\mbox{.}(2015){Feruglio}, {Fiore}, {Carniani},
  {Piconcelli}, {Zappacosta}, {Bongiorno}, {Cicone}, {Maiolino}, {Marconi},
  {Menci}, {Puccetti}, \& {Veilleux}}]{Fergulio+2015}
{Feruglio} C. {et~al.}, 2015, \aap, 583, A99

\bibitem[{{Fischera} \& {Dopita}(2004)}]{Fischera+Dopita2004}
{Fischera} J., {Dopita} M.~A., 2004, \apj, 611, 919

\bibitem[{{Fischera} {et~al}\mbox{.}(2003){Fischera}, {Dopita}, \&
  {Sutherland}}]{Fischera+2003}
{Fischera} J., {Dopita} M.~A., {Sutherland} R.~S., 2003, \apjl, 599, L21

\bibitem[{{Gabor} \& {Bournaud}(2014)}]{Gabor+Bournaud2014}
{Gabor} J.~M., {Bournaud} F., 2014, \mnras, 441, 1615

\bibitem[{{Gaibler} {et~al}\mbox{.}(2012){Gaibler}, {Khochfar}, {Krause}, \&
  {Silk}}]{Gaibler+2012}
{Gaibler} V., {Khochfar} S., {Krause} M., {Silk} J., 2012, \mnras, 425, 438

\bibitem[{{Genzel} {et~al}\mbox{.}(2010){Genzel}, {Tacconi}, {Gracia-Carpio},
  {Sternberg}, {Cooper}, {Shapiro}, {Bolatto}, {Bouch{\'e}}, {Bournaud},
  {Burkert}, {Combes}, {Comerford}, {Cox}, {Davis}, {Schreiber},
  {Garcia-Burillo}, {Lutz}, {Naab}, {Neri}, {Omont}, {Shapley}, \&
  {Weiner}}]{Genzel+2010}
{Genzel} R. {et~al.}, 2010, \mnras, 407, 2091

\bibitem[{{Gnedin} \& {Abel}(2001)}]{Gnedin+Abel2001}
{Gnedin} N.~Y., {Abel} T., 2001, New Astronomy, 6, 437

\bibitem[{{Hopkins} {et~al}\mbox{.}(2011){Hopkins}, {Quataert}, \&
  {Murray}}]{Hopkins+2011}
{Hopkins} P.~F., {Quataert} E., {Murray} N., 2011, \mnras, 417, 950

\bibitem[{{Hopkins} {et~al}\mbox{.}(2007){Hopkins}, {Richards}, \&
  {Hernquist}}]{Hopkins+2007}
{Hopkins} P.~F., {Richards} G.~T., {Hernquist} L., 2007, \apj, 654, 731

\bibitem[{{Hopkins} {et~al}\mbox{.}(2016){Hopkins}, {Torrey},
  {Faucher-Gigu{\`e}re}, {Quataert}, \& {Murray}}]{Hopkins+2016}
{Hopkins} P.~F., {Torrey} P., {Faucher-Gigu{\`e}re} C.-A., {Quataert} E.,
  {Murray} N., 2016, \mnras, 458, 816

\bibitem[{{Hu}(2008)}]{Hu2008}
{Hu} J., 2008, \mnras, 386, 2242

\bibitem[{{Jiang} {et~al}\mbox{.}(2012){Jiang}, {Stone}, \&
  {Davis}}]{Jiang+Stone+Davis2012}
{Jiang} Y.-F., {Stone} J.~M., {Davis} S.~W., 2012, \apjs, 199, 14

\bibitem[{{Khandai} {et~al}\mbox{.}(2015){Khandai}, {Di Matteo}, {Croft},
  {Wilkins}, {Feng}, {Tucker}, {DeGraf}, \& {Liu}}]{MBII}
{Khandai} N., {Di Matteo} T., {Croft} R., {Wilkins} S., {Feng} Y., {Tucker} E.,
  {DeGraf} C., {Liu} M.-S., 2015, \mnras, 450, 1349

\bibitem[{{Kim} {et~al}\mbox{.}(2011){Kim}, {Wise}, {Alvarez}, \&
  {Abel}}]{Kim+2011}
{Kim} J.-h., {Wise} J.~H., {Alvarez} M.~A., {Abel} T., 2011, \apj, 738, 54

\bibitem[{{King}(2003)}]{King2003}
{King} A., 2003, \apjl, 596, L27

\bibitem[{{Kormendy} {et~al}\mbox{.}(2011){Kormendy}, {Bender}, \&
  {Cornell}}]{Kormendy+2011}
{Kormendy} J., {Bender} R., {Cornell} M.~E., 2011, \nat, 469, 374

\bibitem[{{Kormendy} \& {Ho}(2013)}]{Kormendy+Ho2013}
{Kormendy} J., {Ho} L.~C., 2013, \araa, 51, 511

\bibitem[{{Kritsuk} {et~al}\mbox{.}(2011){Kritsuk}, {Norman}, \&
  {Wagner}}]{Kritsuk+2011}
{Kritsuk} A.~G., {Norman} M.~L., {Wagner} R., 2011, \apjl, 727, L20

\bibitem[{{Krumholz} {et~al}\mbox{.}(2011){Krumholz}, {Klein}, \&
  {McKee}}]{Krumholz+Klein+McKee2011}
{Krumholz} M.~R., {Klein} R.~I., {McKee} C.~F., 2011, \apj, 740, 74

\bibitem[{{Krumholz} \& {Matzner}(2009)}]{Krumholz&Matzner2009}
{Krumholz} M.~R., {Matzner} C.~D., 2009, \apj, 703, 1352

\bibitem[{{Krumholz} \& {Thompson}(2012)}]{KT2012}
{Krumholz} M.~R., {Thompson} T.~A., 2012, \apj, 760, 155

\bibitem[{{Krumholz} \& {Thompson}(2013)}]{KT2013}
{Krumholz} M.~R., {Thompson} T.~A., 2013, \mnras, 434, 2329

\bibitem[{{Levermore}(1984)}]{Levermore1984}
{Levermore} C.~D., 1984, \jqsrt, 31, 149

\bibitem[{Lewis \& Austin(2002)}]{Lewis+Austin2002}
Lewis G.~M., Austin P.~H., 2002, Jp4.16 an iterative method for generating
  scaling log-normal simulations

\bibitem[{{Li} \& {Draine}(2001)}]{Li+Draine2001}
{Li} A., {Draine} B.~T., 2001, \apj, 554, 778

\bibitem[{{Magorrian} {et~al}\mbox{.}(1998){Magorrian}, {Tremaine},
  {Richstone}, {Bender}, {Bower}, {Dressler}, {Faber}, {Gebhardt}, {Green},
  {Grillmair}, {Kormendy}, \& {Lauer}}]{Magorrian1998}
{Magorrian} J. {et~al.}, 1998, \aj, 115, 2285

\bibitem[{{McKee} \& {Ostriker}(2007)}]{McKee+Ostriker2007}
{McKee} C.~F., {Ostriker} E.~C., 2007, \araa, 45, 565

\bibitem[{{Navarro} {et~al}\mbox{.}(1996){Navarro}, {Frenk}, \&
  {White}}]{NFW1996}
{Navarro} J.~F., {Frenk} C.~S., {White} S.~D.~M., 1996, \apj, 462, 563

\bibitem[{{Nayakshin} \& {Zubovas}(2012)}]{Nayakshin+Zubovas2012}
{Nayakshin} S., {Zubovas} K., 2012, \mnras, 427, 372

\bibitem[{{Novak} {et~al}\mbox{.}(2012){Novak}, {Ostriker}, \&
  {Ciotti}}]{Novak+2012}
{Novak} G.~S., {Ostriker} J.~P., {Ciotti} L., 2012, \mnras, 427, 2734

\bibitem[{{Ostriker} {et~al}\mbox{.}(2010){Ostriker}, {Choi}, {Ciotti},
  {Novak}, \& {Proga}}]{Ostriker+2010}
{Ostriker} J.~P., {Choi} E., {Ciotti} L., {Novak} G.~S., {Proga} D., 2010,
  \apj, 722, 642

\bibitem[{{Pawlik} \& {Schaye}(2011)}]{Pawlik+Schaye2011}
{Pawlik} A.~H., {Schaye} J., 2011, \mnras, 412, 1943

\bibitem[{{Petkova} \& {Springel}(2009)}]{Petkova+Springel2009}
{Petkova} M., {Springel} V., 2009, \mnras, 396, 1383

\bibitem[{{Proga} {et~al}\mbox{.}(2000){Proga}, {Stone}, \&
  {Kallman}}]{Proga+2000}
{Proga} D., {Stone} J.~M., {Kallman} T.~R., 2000, \apj, 543, 686

\bibitem[{{Roos} {et~al}\mbox{.}(2015){Roos}, {Juneau}, {Bournaud}, \&
  {Gabor}}]{roosetal}
{Roos} O., {Juneau} S., {Bournaud} F., {Gabor} J.~M., 2015, \apj, 800, 19

\bibitem[{{Rosdahl} {et~al}\mbox{.}(2013){Rosdahl}, {Blaizot}, {Aubert},
  {Stranex}, \& {Teyssier}}]{Rosdahl+2013}
{Rosdahl} J., {Blaizot} J., {Aubert} D., {Stranex} T., {Teyssier} R., 2013,
  \mnras, 436, 2188

\bibitem[{{Rosdahl} \& {Teyssier}(2015{\natexlab{a}})}]{RT2015}
{Rosdahl} J., {Teyssier} R., 2015{\natexlab{a}}, \mnras, 449, 4380

\bibitem[{{Rosdahl} \& {Teyssier}(2015{\natexlab{b}})}]{Rosdahl+2015}
{Rosdahl} J., {Teyssier} R., 2015{\natexlab{b}}, \mnras, 449, 4380

\bibitem[{{Ro{\v s}kar} {et~al}\mbox{.}(2014){Ro{\v s}kar}, {Teyssier},
  {Agertz}, {Wetzstein}, \& {Moore}}]{Roskar+2014}
{Ro{\v s}kar} R., {Teyssier} R., {Agertz} O., {Wetzstein} M., {Moore} B., 2014,
  \mnras, 444, 2837

\bibitem[{{Sazonov} {et~al}\mbox{.}(2004){Sazonov}, {Ostriker}, \&
  {Sunyaev}}]{Sazonov+2004}
{Sazonov} S.~Y., {Ostriker} J.~P., {Sunyaev} R.~A., 2004, \mnras, 347, 144

\bibitem[{{Schaye} {et~al}\mbox{.}(2015){Schaye}, {Crain}, {Bower}, {Furlong},
  {Schaller}, {Theuns}, {Dalla Vecchia}, {Frenk}, {McCarthy}, {Helly},
  {Jenkins}, {Rosas-Guevara}, {White}, {Baes}, {Booth}, {Camps}, {Navarro},
  {Qu}, {Rahmati}, {Sawala}, {Thomas}, \& {Trayford}}]{Eagle}
{Schaye} J. {et~al.}, 2015, \mnras, 446, 521

\bibitem[{{Schechter}(1976)}]{Schechter1976}
{Schechter} P., 1976, \apj, 203, 297

\bibitem[{{Semenov} {et~al}\mbox{.}(2003){Semenov}, {Henning}, {Helling},
  {Ilgner}, \& {Sedlmayr}}]{Semenov+2003}
{Semenov} D., {Henning} T., {Helling} C., {Ilgner} M., {Sedlmayr} E., 2003,
  \aap, 410, 611

\bibitem[{{Shakura} \& {Sunyaev}(1973)}]{Shakura&Sunyaev1973}
{Shakura} N.~I., {Sunyaev} R.~A., 1973, \aap, 24, 337

\bibitem[{{Sijacki} {et~al}\mbox{.}(2007){Sijacki}, {Springel}, {Di Matteo}, \&
  {Hernquist}}]{Sijacki+2007}
{Sijacki} D., {Springel} V., {Di Matteo} T., {Hernquist} L., 2007, \mnras, 380,
  877

\bibitem[{{Silk} \& {Rees}(1998)}]{Silk+Rees1998}
{Silk} J., {Rees} M.~J., 1998, \aap, 331, L1

\bibitem[{{Skinner} \& {Ostriker}(2013)}]{Skinner+OstrikerE2013}
{Skinner} M.~A., {Ostriker} E.~C., 2013, \apjs, 206, 21

\bibitem[{{Sutherland} \& {Bicknell}(2007)}]{Sutherland+Bicknell2007}
{Sutherland} R.~S., {Bicknell} G.~V., 2007, \apjs, 173, 37

\bibitem[{{Tacconi} {et~al}\mbox{.}(2010){Tacconi}, {Genzel}, {Neri}, {Cox},
  {Cooper}, {Shapiro}, {Bolatto}, {Bouch{\'e}}, {Bournaud}, {Burkert},
  {Combes}, {Comerford}, {Davis}, {Schreiber}, {Garcia-Burillo},
  {Gracia-Carpio}, {Lutz}, {Naab}, {Omont}, {Shapley}, {Sternberg}, \&
  {Weiner}}]{Tacconi+2010}
{Tacconi} L.~J. {et~al.}, 2010, \nat, 463, 781

\bibitem[{{Teyssier}(2002)}]{Teyssier2002}
{Teyssier} R., 2002, \aap, 385, 337

\bibitem[{{Tombesi} {et~al}\mbox{.}(2015){Tombesi}, {Mel{\'e}ndez}, {Veilleux},
  {Reeves}, {Gonz{\'a}lez-Alfonso}, \& {Reynolds}}]{Tombesi+2015}
{Tombesi} F., {Mel{\'e}ndez} M., {Veilleux} S., {Reeves} J.~N.,
  {Gonz{\'a}lez-Alfonso} E., {Reynolds} C.~S., 2015, \nat, 519, 436

\bibitem[{{Toro} {et~al}\mbox{.}(1994){Toro}, {Spruce}, \&
  {Speares}}]{Toro+1994}
{Toro} E.~F., {Spruce} M., {Speares} W., 1994, Shock Waves, 4, 25

\bibitem[{{Tsang} \& {Milosavljevi{\'c}}(2015)}]{Tsang+2015}
{Tsang} B.~T.-H., {Milosavljevi{\'c}} M., 2015, \mnras, 453, 1108

\bibitem[{{Verner} {et~al}\mbox{.}(1996){Verner}, {Ferland}, {Korista}, \&
  {Yakovlev}}]{Verner+1996}
{Verner} D.~A., {Ferland} G.~J., {Korista} K.~T., {Yakovlev} D.~G., 1996, \apj,
  465, 487

\bibitem[{{Vogelsberger} {et~al}\mbox{.}(2014){Vogelsberger}, {Genel},
  {Springel}, {Torrey}, {Sijacki}, {Xu}, {Snyder}, {Nelson}, \&
  {Hernquist}}]{Illustris}
{Vogelsberger} M. {et~al.}, 2014, \mnras, 444, 1518

\bibitem[{{Wagner} \& {Bicknell}(2011)}]{Wagner+Bicknell2011}
{Wagner} A.~Y., {Bicknell} G.~V., 2011, \apj, 728, 29

\bibitem[{{Wagner} {et~al}\mbox{.}(2013){Wagner}, {Umemura}, \&
  {Bicknell}}]{Wagner+2013}
{Wagner} A.~Y., {Umemura} M., {Bicknell} G.~V., 2013, \apjl, 763, L18

\bibitem[{{Warhaft}(2000)}]{Warhaft2000}
{Warhaft} Z., 2000, Annual Review of Fluid Mechanics, 32, 203

\bibitem[{{Wise} \& {Abel}(2011)}]{Wise+Abel2011}
{Wise} J.~H., {Abel} T., 2011, \mnras, 414, 3458

\bibitem[{{Zubovas} \& {King}(2012)}]{Zubovas&King2012}
{Zubovas} K., {King} A., 2012, \apjl, 745, L34

\bibitem[{{Zubovas} \& {Nayakshin}(2014)}]{Zubovas&Nayakshin2014}
{Zubovas} K., {Nayakshin} S., 2014, \mnras, 440, 2625

\bibitem[{{Zubovas} {et~al}\mbox{.}(2013){Zubovas}, {Nayakshin}, {King}, \&
  {Wilkinson}}]{zubovasetal13}
{Zubovas} K., {Nayakshin} S., {King} A., {Wilkinson} M., 2013, \mnras, 433,
  3079

\end{thebibliography}

\appendix

\section{Effect of different speed of light approximations}
\label{app:SOL}
\begin{figure}
\includegraphics[width=\columnwidth]{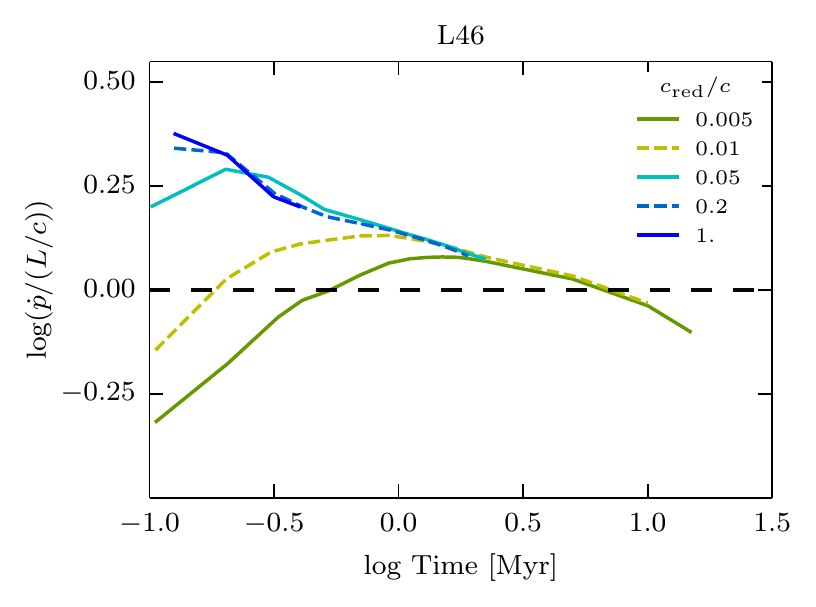}
\caption{Convergence test of the mechanical advantage for the lower resolution
\Le{46}\_\medC\_\textit{med$\rho_{\rm Q}$} simulation using different values
for the reduced speed of light $c_{\rm red}$, whose default value is $0.2\,c$
in our simulations. Convergence occurs for $c_{\rm red} \ge 0.2\,c$.}
\label{fig:PropSOL}
\end{figure}

With the explicit M1 scheme that we use for the transport of radiation between grid
cells, the RHD timestep is limited by the speed of light, since the solver
breaks down if radiation is allowed to travel more than one cell width in one
timestep. We thus use the reduced speed of light approximation
\citep{Gnedin+Abel2001}, with a fiducial light speed of $c_{\rm red} =0.2 \, c$
in this work.

To understand how much our reduced speed of light affects our results, we have
run lower-resolution equivalents to \Le{46}\_\medC\_\textit{med$\rho_{\rm Q}$}
with a spatial resolution of $\Delta x = 11.6$~pc (as opposed to fiducial
resolution of $\Delta x = 5.8$~pc used in the rest of the paper), and where we
adopt different values for the reduced speed of light $c_{\rm red}$.  The
evolution of the mechanical advantage, shown in Fig.~\ref{fig:PropSOL},
indicates that the choice of $c_{\rm red}$ has a significant effect on the
mechanical advantage, especially at the beginning of the simulation.  In fact,
the mechanical advantage can increase by an order of magnitude, at early times
($0.1\, \rm Myr$), if the reduced speed of light fraction is increased from
$c_{\rm red}=0.005$ to unity. Yet, this increase is very limited when changing
from $c_{\rm red} = 0.2\,c$ (as in our study) to $c$.  After $\sim$~2~Myr, when
the radiation has managed to carve a hole through which it escapes, all runs
converge towards the same momentum input rate.

We conclude that our results regarding the mechanical
advantage are not significantly altered by the reduced speed of light to $0.2\,c$ and that
our results are thus well converged.  In a forthcoming paper (Bieri et al, in
prep.), we will discuss in more detail the measured reduction factor and its
dependence on the reduced speed of light.

\end{document}